\documentclass[sigconf]{acmart}

\usepackage{url}
\usepackage{graphicx}
\usepackage{amsmath}
\usepackage{float}          
\usepackage{multirow}       
\usepackage{subfig}
\usepackage{algorithm}
\usepackage{algorithmic}

\newcommand{\PP}[1]{
\noindent{\bf #1}
}
\newcommand{\argmin}{\operatornamewithlimits{\mathstrut argmin}}

\AtBeginDocument{%
  \providecommand\BibTeX{{%
    \normalfont B\kern-0.5em{\scshape i\kern-0.25em b}\kern-0.8em\TeX}}}

\setcopyright{acmcopyright}
\copyrightyear{2021}
\acmYear{2021}
\setcopyright{acmlicensed}\acmConference[ACSAC '21]{Annual Computer Security Applications Conference}{December 6--10, 2021}{Virtual Event, USA}
\acmBooktitle{Annual Computer Security Applications Conference (ACSAC '21), December 6--10, 2021, Virtual Event, USA}
\acmPrice{15.00}
\acmDOI{10.1145/3485832.3485912}
\acmISBN{978-1-4503-8579-4/21/12}

\begin{document}

\title{Detecting Audio Adversarial Examples with Logit Noising}

\author{Namgyu Park}
\email{namgyu.park@postech.ac.kr}
\affiliation{%
  \institution{POSTECH}
  \city{Pohang}
  \country{South Korea}
}

\author{Sangwoo Ji}
\email{sangwooji@postech.ac.kr}
\affiliation{%
  \institution{POSTECH}
  \city{Pohang}
  \country{South Korea}
}

\author{Jong Kim}
\email{jkim@postech.ac.kr}
\affiliation{%
  \institution{POSTECH}
  \city{Pohang}
  \country{South Korea}
}

\renewcommand{\shortauthors}{Park, et al.}

\begin{abstract}
Automatic speech recognition (ASR) systems are vulnerable to audio adversarial examples that attempt to deceive ASR systems by adding perturbations to benign speech signals. 
Although an adversarial example and the original benign wave are indistinguishable to humans, the former is transcribed as a malicious target sentence by ASR systems.
Several methods have been proposed to generate audio adversarial examples and feed them directly into the ASR system (over-line). 
Furthermore, many researchers have demonstrated the feasibility of robust physical audio adversarial examples (over-air).
To defend against the attacks, several studies have been proposed.
However, deploying them in a real-world situation is difficult because of accuracy drop or time overhead.

In this paper, we propose a novel method to detect audio adversarial examples by adding noise to the logits before feeding them into the decoder of the ASR. 
We show that carefully selected noise can significantly impact the transcription results of the audio adversarial examples, whereas it has minimal impact on the transcription results of benign audio waves.
Based on this characteristic, we detect audio adversarial examples by comparing the transcription altered by logit noising with its original transcription.
The proposed method can be easily applied to ASR systems without any structural changes or additional training.
The experimental results show that the proposed method is robust to over-line audio adversarial examples as well as over-air audio adversarial examples compared with state-of-the-art detection methods.

\end{abstract}

\begin{CCSXML}
<ccs2012>
<concept>
<concept_id>10002978</concept_id>
<concept_desc>Security and privacy</concept_desc>
<concept_significance>500</concept_significance>
</concept>
<concept>
<concept_id>10010147.10010178.10010179.10010183</concept_id>
<concept_desc>Computing methodologies~Speech recognition</concept_desc>
<concept_significance>500</concept_significance>
</concept>
</ccs2012>
\end{CCSXML}

\ccsdesc[500]{Security and privacy}
\ccsdesc[500]{Computing methodologies~Speech recognition}

\keywords{automatic speech recognition system, audio adversarial examples, logits, adversarial example detection, over-line \& over-air attack}

\maketitle

\section{Introduction}

With the development of deep learning techniques, automatic speech recognition (ASR) systems have been developed rapidly. 
Owing to the development of ASR systems, intelligent voice assistants (IVA), such as Google Assistant~\cite{Assistant}, Amazon Alexa~\cite{Alexa}, Apple Siri~\cite{Siri}, and Microsoft Cortana~\cite{Cortana} are widely used in human-machine interfaces. 
Through voice command, people can get their smartphones or smart speakers to schedule their personal appointments, open doors, and send text messages.  

\begin{figure}[t]
\begin{center}
  \includegraphics[width=.95\linewidth]{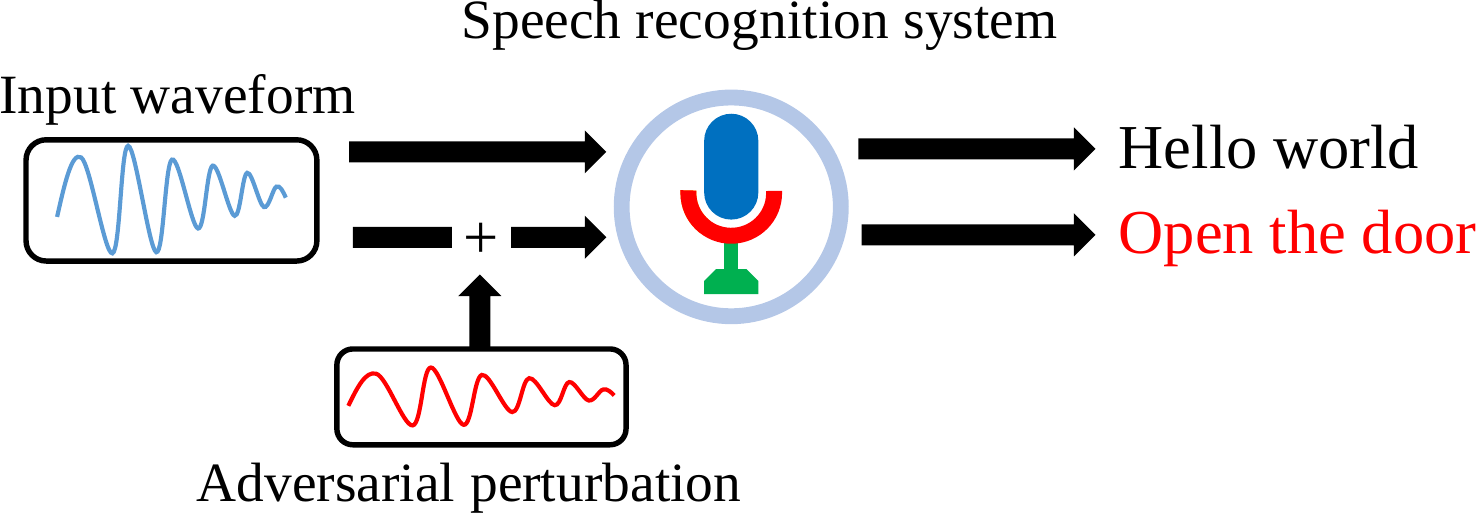}
\end{center}
\caption{The overview of generating audio adversarial examples. Attackers add perturbation to benign signal. The generated audio adversarial example sounds like "hello world", but it is recognized as "open the door" in a target ASR model.}
\label{fig:aae}
\end{figure}

However, ASR systems are vulnerable to audio adversarial examples~\cite{abdullah2020sok}, which are malicious audio signals made by adding small perturbations to benign speech signals to fool deep neural networks (DNNs).
Humans cannot distinguish these examples from the original contents of the speech, although the signals are slightly noisy.
However, they are recognized as malicious content by ASR systems (Figure~\ref{fig:aae}).
Many researchers have proposed methods to generate adversarial examples to attack DNN-based ASR systems~\cite{hannun2014deep,povey2011kaldi,lingvo}.
Most generated adversarial examples are directly fed into the ASR system over the line~\cite{C_W,hiromu,weighted_aaai20,blackboxSPW19,schonherr2018adversarial}.
Furthermore, researchers have demonstrated the feasibility of attacking ASR systems using over-air adversarial audio signals known as physical audio adversarial examples~\cite{hiromu,tao2020Metamorph,yaoqinicml,yuan2018commandersong,20imperio,develsecurity@2020}, which are made by considering real environmental audio noises when the audio signals are played over the air. 
These attacks can pose a severe risk to IVAs and control systems. 
For example, an attacker can conceal an attack command in a piece of music or a voice message on the radio, either to get money sent to a specific account or to keep the navigation in search of the wrong location. 

\begin{figure*}[t]
\begin{center}
  \includegraphics[width=.95\linewidth]{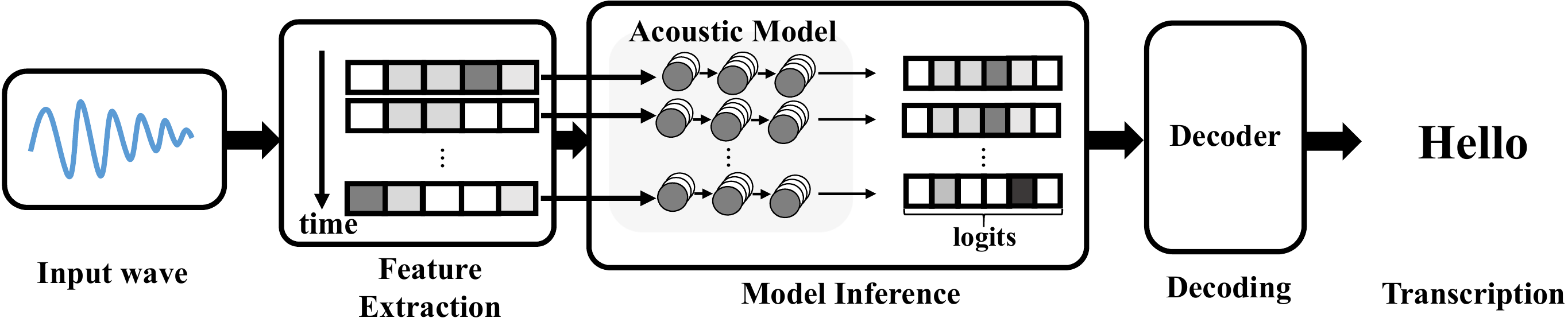}
\end{center}
\caption{The overview of DNN-based ASR process. There are total three phases: feature extraction, model inference, and decoding}
\label{fig:asr}
\end{figure*}

Several researchers have proposed methods to defend against the attacks.
A line of work has tried to train a robust ASR model~\cite{adv_training1,adv_training2,adv_training3}; however, these methods suffer a significant accuracy drop~\cite{abdullah2020sok,develsecurity@2020}. 
Meanwhile, methods to detect adversarial examples have been proposed.
Kwon \textsl{et al.}~\cite{Kwon19ccsw} proposed a detection method using audio preprocessing such as low-pass filters or high-pass filters, and Zhuolin \textsl{et al.}~\cite{yang2018characterizing} proposed a detection method using audio splitting. 
Although these methods seem to be effective for over-line audio adversarial examples, these research did not consider over-air audio adversarial examples. 
Recently, a new method based on signal reverberations~\cite{MM20detection} was proposed to detect robust physical audio adversarial examples.
However, this method requires multiple wave inferences to evaluate whether the input wave is an audio adversarial example or not; therefore, it has a large time overhead.

In this paper, we propose logit noising, a novel method to detect an audio adversarial example.
We add noise to logits, which is the output of an acoustic model in ASR system~(Figure~\ref{fig:asr}).
The method is based on the observation that the logits of benign audio waves and audio adversarial examples have different distributions.
Therefore, the transcriptions of benign audio waves are rarely changed following logit noising, whereas those of adversarial examples are easily altered.
We then distinguish adversarial examples from benign waves by comparing the extent of transcription distortions arising from logit noising.
Through extensive experiments, we show that the proposed method can achieve 100\% detection accuracy on over-line audio adversarial examples and over 95.0\% accuracy on over-air attacks.
The proposed method is easily applicable to existing trained models because it requires no structural change or no additional training procedure.
Moreover, as the method requires no additional inference for an input, it reduces the time overhead compared with state-of-the-art detection methods~\cite{Kwon19ccsw,yang2018characterizing,MM20detection}.

Our main contributions are as follows:
\begin{itemize}
    \item We show that logits of benign audio waves and adversarial examples have different characteristics. 
    We then conduct an analysis to determine the threshold for distinguishing adversarial examples.
    
    \item We propose logit noising, a novel method to detect audio adversarial examples by adding noise to the intermediate results (logits) of ASR systems. 
    We demonstrate that the proposed method can effectively detect both over-air and over-line adversarial examples.
    
    \item We show that the proposed method can be easily deployed in real-world ASR systems because of its high compatibility (no structural change or retraining) and high efficiency (less overhead).
\end{itemize}

The remainder of this paper is organized as follows. 
In Section 2, we present background information on ASR systems.
In Section 3, we present the related research on audio adversarial attacks and defense methods.
The proposed detection method and experimental evaluation results are presented in Section 4 and 5, respectively.
In Section 6, we provide the discussion.
In Section 7, we provide our conclusions.
\section{Background: Automatic Speech Recognition}
\label{sec:back}

ASR is a process in which a system interprets human speech and converts the content into text data.
In this work, we experiment on the open-source DNN-based ASR framework DeepSpeech~\cite{hannun2014deep}, which has been the target of many audio adversarial examples~\cite{C_W,hiromu,tao2020Metamorph,weighted_aaai20}, to demonstrate our proposed method. Figure~\ref{fig:asr} is an overview of this system. The ASR architecture consists of three major components for audio transcription, namely, feature extraction, model inference, and decoding.\\

\PP{Feature extraction.}
In this phase, the input wave is converted to a mel-frequency cepstrum coefficient (MFCC)~\cite{MFCC} based on human hearing perceptions. 
First, the input wave is split into short time frames (20 $\sim$ 40 ms) with overlapping.
For each frame, a fast Fourier transform~\cite{FFT} is used to generate the frequency domain data.
However, the linearly spaced result cannot characterize human perception effectively because humans perceive  frequency non-linearly.
To approximate the human perception, the frequency domain data are converted to mel-scaled data using mel-filter.
Subsequently, MFCC feature vectors are obtained by taking the discrete cosine transform~\cite{DCT} of the log of mel-scaled data.\\

\PP{Model inference.} 
The extracted MFCC features are passed to a DNN acoustic model. The target system uses a recurrent neural network (RNN) as the DNN acoustic model. Generally, an RNN-based acoustic model accepts a variable-sized MFCC vector as input. 
The acoustic model then outputs the sequence of character likelihood. 
In this paper, we call one character likelihood as logits (Figure~\ref{fig:asr}: Model Inference).
\\

\PP{Decoding.} 
The logits sequence is transcribed to character-level transcription using beam-search decoding. 
The beam-search decoding functions by taking the logits and looking for the most likely text sequence according to the logits sequence.
Finally, the character-level transcription is passed through a language model to be properly transcribed.
The language model outputs the final transcription based on word and semantic proximity using the N-gram model.
\section{Related Work}

\subsection{Adversarial Attacks in ASR}
\label{sec:attack}
Adversarial examples are maliciously crafted audio waves that are intended to cause mistranscriptions by the ASR systems~\cite{C_W,yuan2018commandersong,hiromu}.
Most attacks are aimed at producing the targeted malicious sentences; they are a severe security threat~\cite{tao2020Metamorph,advpulse}.
For example, an attacker may open a door by submitting adversarial examples that are perceived as ``hello world'' to the human hearing system (Figure~\ref{fig:aae}).

To generate such examples, an attacker is required to possess a certain knowledge of the target ASR system.
A white-box attack is developed under the assumption that an attacker has full knowledge of the target ASR system including the DNN structure, trained parameters, and preprocessing method.
The attacker then generates an example by adjusting the example using gradients through the full ASR process.
A white-box attack is the best scenario for an attacker, as it can generate the most powerful adversarial examples.
A black-box attack is developed under the assumption that an attacker cannot access the internal parameters of the ASR system.
Therefore, the attacker must generate an adversarial example using only the transcription result from the target system. 

\begin{figure}%
    \centering
    \subfloat[Over-line attack\label{fig:overline}]{{\includegraphics[width=7cm]{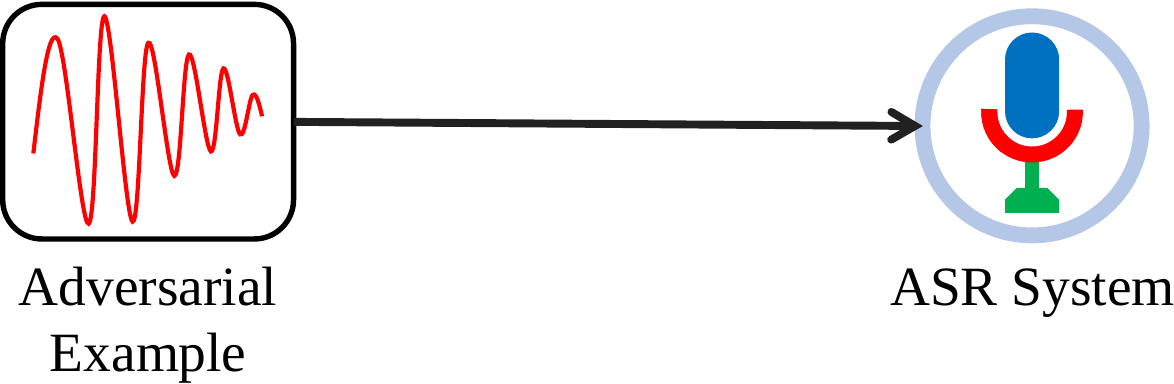} }}%
    \qquad
    \subfloat[Over-air attack\label{fig:overair}]{{\includegraphics[width=7cm]{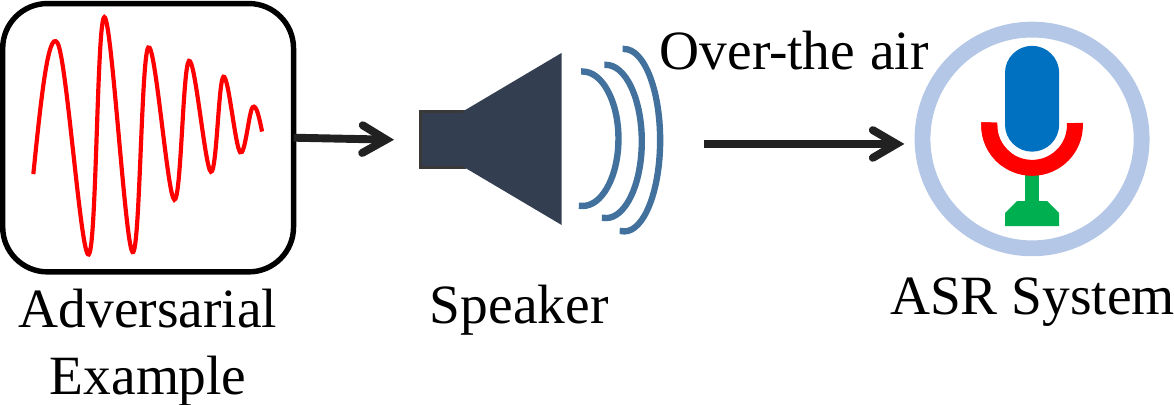} }}%
    \caption{Two attack types for audio adversarial examples.}%
\end{figure}

In this study, we aim to detect two types of audio adversarial attacks: over-line and over-air attacks.
The first is a type of attack that assumes that the generated adversarial audio wave is directly passed to the ASR system (Figure~\ref{fig:overline}). 
Then, the adversarial wave is mistranscribed by the ASR system without any interference.
On the other hand, in an over-air attack (Figure~\ref{fig:overair}), it is assumed that the adversarial audio wave is aired through the speaker to the target ASR system.
During recording, the wave is affected by ambient noise, the room environment, and the capability of the recording device.
Therefore, over-air attacks are designed to maintain the adversarial effect on waves in different environments.
We provide detailed examples of both types of attacks below.\\

\subsubsection{Over-line attack}~

\PP{C\&W.}~\cite{C_W} Carlini \& Wanger proposed a white-box targeted attack. 
The authors showed that any source audio can be mistranslated to any attacker's intended phrase through small perturbation. 
In this attack, an attacker selects a target phrase and uses a connectionist temporal classification (CTC) loss~\cite{CTC} function as an objective function. 
The CTC-loss sums up the probability of possible alignments of input to target, producing a loss value which is differentiable with respect to each input. From this, an attacker can continually update audio adversarial examples through gradient-descent. 

$$
\argmin_{\delta} \; dB_{x}(\delta) + c \cdot L_{ctc}(ASR(x + \delta),t)
\label{eq:fgsmattack1}    
$$
where $t$ is the target phrase and $c$ is the importance of the adversarial part. 
An attacker can achieve a human-imperceptible but machine-effective adversarial perturbation by optimizing this equation.\\

\PP{Taori.}~\cite{blackboxSPW19} 
Taori \textit{et al.} proposed a black-box attack. 
They introduced a momentum mutation inspired by the momentum update for gradient descent. With momentum update, the mutation probability changes according to the following exponentially weighted moving average update:

$$
p_{new} = \alpha \cdot p_{old} + {\frac{\beta}{|curr\_Score - prev\_Score|}}
$$
where $Score$ is CTC-loss. For each iteration, they computed the $Score$ and updated the new mutation probability, thereby creating optimal audio adversarial examples.\\

\subsubsection{Over-air attack}~

\PP{Hiromu.}~\cite{hiromu} 
Hiromu \textit{et al.} presented an advanced white-box targeted adversarial attack. 
They investigated the feasibility of attacks using audio adversarial examples in the physical world, that is, attacks whereby a speaker makes a sound over the air and a microphone receives it. To generate over-air audio adversarial examples, the authors suggested the optimization using expectation over transformation (EoT)~\cite{EOT18ICML}:

$$
\argmin_{\delta} \; \mathbb{E}_{h \sim \mathcal{H}}[dB_{x}(\delta) + c \cdot L_{ctc}(ASR(x' + \delta),t)] 
$$
$$
\text{where} \; x' = Conv_h(x+BPF_{1000 \sim 4000Hz} (\delta)).
$$
$BPF$ means band-pass filter, $Conv$ means convolution, and $\mathcal{H}$ represents a set of collected impulse responses. 
They showed the feasibility of such attacks considering various reverberations and ambient noises.

In general, over-air attacks are more robust than over-line attacks because they are generated based on various external factors such as background noise and device internal noise.

\begin{figure*}[t]
\begin{center}
  \includegraphics[width=.95\linewidth]{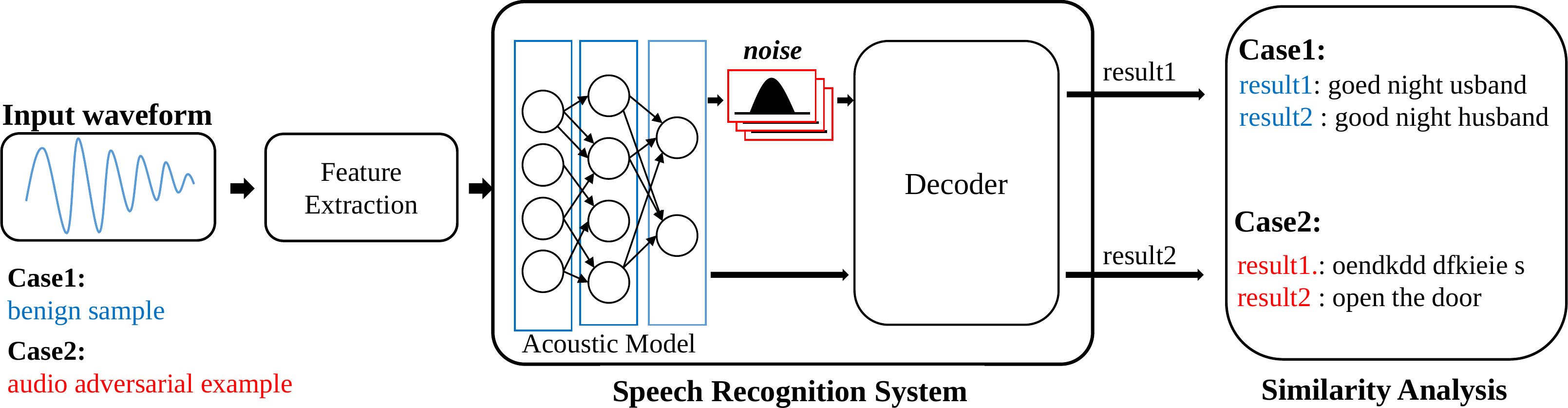}
\end{center}
\caption{Logit noising architecture}
\label{fig:LNArchitecture}
\end{figure*}

\subsection{Detection Methods}
Defenses against audio adversarial examples have been extensively researched.
To create a robust speech recognition system, many researchers have adopted adversarial training techniques~\cite{adv_training1,adv_training2,adv_training3}, which have achieved excellent performance in image classification tasks~\cite{madry2017towards}.
Adversarial training methods entail a robust model using adversarial examples during training procedure.
In audio domain, however, adversarial training suffers a noticeable accuracy drop on benign test dataset~\cite{advertraining@sp21,abdullah2020sok}.
Meanwhile, detection methods that do not affect the accuracy on the benign data have been proposed.\\

\PP{Audio random filter.}
Kwon \textsl{et al.}~\cite{Kwon19ccsw} proposed input audio modification using random filters such as high-pass, low-pass, or notch filters. 
They argued that the effect of adversarial perturbation, which is widely spread in frequency domain, can be reduced when these filters are applied to audio adversarial examples.
Their experimental results showed that audio adversarial examples have different result ($ASR(x^{adv})$) when they are modified using random filters ($ASR(RandFilter(x^{adv}))$), whereas benign audio waves have similar results ($ASR(x)$) even when they are modified in the same way ($ASR(RandFilter(x))$).
Using this characteristic, the method distinguishes adversarial examples from benign audio waves by comparing the ASR results with and without filtering.
The method, however, has difficulty of preventing over-air attacks because many over-air attacks consider the random filter scheme.\\

\PP{Temporal dependency.}
Zhuolin \textsl{et al.}~\cite{yang2018characterizing} showed that audio sequences have explicit temporal dependency (e.g., correlations in consecutive waveform segments).
To detect audio adversarial examples, they showed that adversarial effects are not effective when the adversarial audio is cut.
Given an audio input, they selected its first $k$ portion (the prefix of length
$k$) as the input for ASR to obtain transcribed results as $ASR(x)_{\{k\}}$, and inserted the whole input into the ASR and selected the prefix of length $k$ of the transcribed result as $ASR(x)_{\{whole,k\}}$, which has the same length as $ASR(x)_{\{k\}}$.
Then, they compared the similarity between both results in terms of temporal dependency distance.
The distance is small for benign audio waves and large for audio adversarial examples.
However, the method does not consider over-air attacks. In addition, its average false positive rate is unsatisfactory based on a specific cutting ratio.\\

\PP{Audio reverberation.}
Xia \textsl{et al.}~\cite{MM20detection} proposed a robust detection method based on an audio reverberation technique. They pioneered the study of defense against robust over-air adversarial examples and discovered that audio adversarial examples are prone to overfitting and continuity. 
The overfitting is broken by room impulse response(RIR)-convolution method ($\mathcal{H}$) into the input wave ($ASR(x* \mathcal{H})$). The RIR-convolution method can overcome the overfitting problem because it significantly modifies the audio waveform.
To break the continuity, they found silent parts of the audio using voice activated detector (VAD) and inserted a Gaussian noise at these parts ($VAD_{silent}(\mathcal{N})$).
Finally, they generated $n$ results with $n$ different RIR-convolutions and padded each silent part using random noise ($ASR(x* \mathcal{H} + VAD_{silent}(\mathcal{N}))$).
They detected adversarial examples by calculating similarity scores of each of $n$ transcription results with reverberation effect and that of the original audio wave.
This method is somewhat effective at detecting over-air audio adversarial examples since various convoluted reverberations make the adversarial effect weaken. 
However, the method has a large time overhead because multiple inferences of generated waves are required.
\section{Proposed Method}

We propose logit noising, a novel detection method that can distinguish audio adversarial examples from benign audio waves. 
We aim to defend against both over-line attacks~\cite{C_W,weighted_aaai20,blackboxSPW19} and over-air attacks~\cite{hiromu,tao2020Metamorph} that have been launched on various settings (white-box or black-box).
All the attacks considered are targeted ones aimed at producing a specific transcription.
Benign audio waves have different logit value gap distributions from audio adversarial examples. 
Therefore, we focus on this logit value gap between them and design the proposed detection method to detect attacks by leveraging the difference in the logit value gap distributions.

The proposed system accepts input audio signals in the same way as the existing system and generates two results for similarity comparison. 
One is the transcription result with original logits of input wave signals, and the other is the agitated transcription result that is produced by adding a certain noise to each logit according to a given noise distribution before feeding the logits to the decoder of ASR. 
Then we compare the similarity of both results. Figure~\ref{fig:LNArchitecture} is an overview of the proposed detection method. 

\subsection{Difference in Logit Value Gap Distribution}
\label{sec:logitsdiff}
As explained in Section~\ref{sec:back}, an input audio wave is converted to a sequence of MFCC feature vectors and passed to the DNN acoustic model (AM). The DNN AM produces a sequence of outputs corresponding to the input sequence. An output is called logits, each element of which represents the likelihood value of one character. Then the decoder (DE) uses the logits to produce a speech transcription. 

\begin{figure}%
    \centering
    \subfloat[1-2 logit-gap distribution\label{fig:1-2gap}]{{\includegraphics[width=7cm]{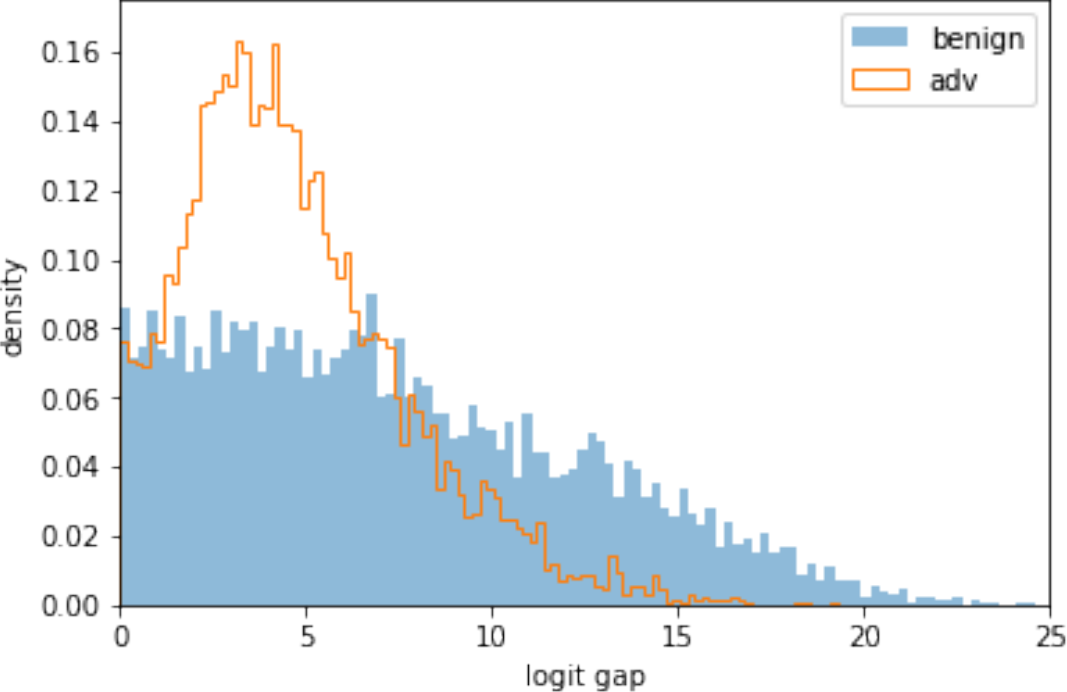} }}%
    \qquad
    \subfloat[1-3 logit-gap distribution\label{fig:1-3gap}]{{\includegraphics[width=7cm]{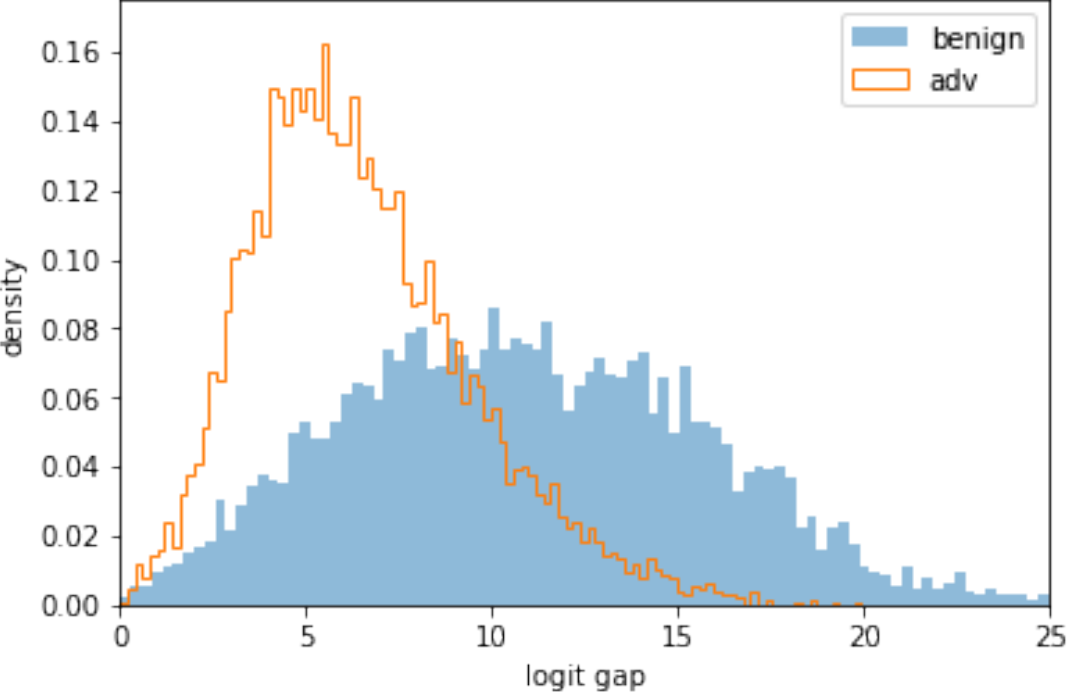} }}%
    \caption{Distribution of the gap between the largest logit value and the second, third largest logit value of 50 benign waves \& adversarial examples.}%
    \label{fig:5}
\end{figure}

We first carefully observe the distribution of the difference between the largest logit value and the remaining $k$-th logit values (1-$k$ logit-gap) in the benign audio waves and audio adversarial examples.
Figure~\ref{fig:1-2gap} shows the 1-2 logit-gap, and Figure~\ref{fig:1-3gap} shows the 1-3 logit-gap of 50 benign audio waves and 50 adversarial examples. 
The x- and y-axes represent the gap and the density of input samples corresponding to the gap, respectively. 
In the figure, the gap of adversarial examples is small and densely distributed with the center at 4 $\sim$ 6, while the gap of benign audio waves is somewhat large and widely distributed for upto 10. 
The difference between benign samples and audio adversarial examples becomes more distinct as the value of $k$ increases.

We conjecture that the audio adversarial examples have a densely distributed and smaller gap because they have to maintain a benign sound to human ears while producing the target phrase.
% by ASR's DE. 
On the other hand, the benign wave samples do not have such restrictions; therefore, the gap is widely distributed.

\subsection{Logit Noising Architecture}
\label{sec:logitsnoise}
To prevent the malicious transcription from being generated in the decoding step for audio adversarial examples, we use the logit noising strategy.
As mentioned before, compared to benign audio waves, there is a smaller gap between the value of the largest logit and the values of the other logits of the audio adversarial examples.
Hence, adding a certain noise to logits can thwart audio adversarial examples by corrupting only the adversarial examples. 
As we add noise to each logit, the order of logit values can be inverted.
If inverted, the decoding step may produce a different transcription from the original transcription. 
If we choose a proper noise level with minimal effect on benign audio and significant effect on adversarial examples, the transcription of the benign waves will remain unchanged, whereas that of the adversarial examples will be changed.

\begin{figure*}[ht]
\begin{center}
  \includegraphics[width=.95\linewidth]{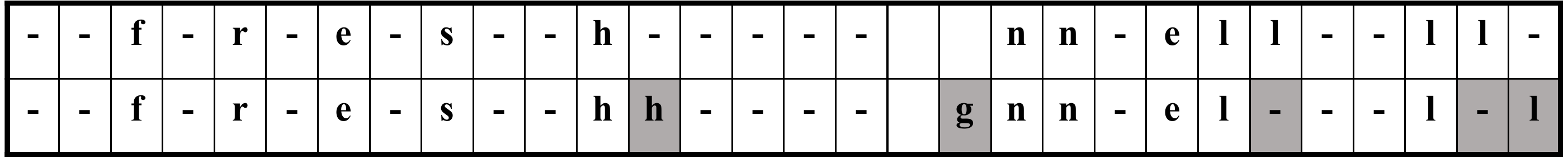}
\end{center}

\caption{Characters of the largest logits after performing the logit noising: upper line shows a benign audio logit argmax result and below line is the same benign audio noised-logits argmax result. Each line is transcribed to "fresh nell" \& "fresh gnell" after passing through decoder.}

\label{fig:NaiveLN}
\end{figure*}

\begin{figure*}[h]
\begin{center}
  \includegraphics[width=.95\linewidth]{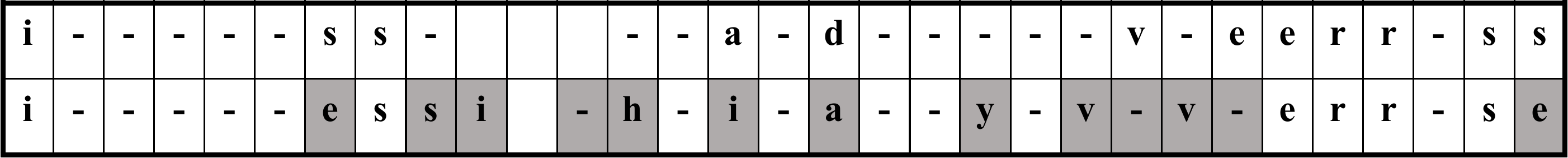}
\end{center}
\caption{Characters of the largest logit after performing the logit noising: upper line shows an audio adversarial example logit argmax result and below line is the same audio adversarial example noised-logits argmax result. Each line is transcribed to " is  advers" \& "iesi hiayof vverse" after passing through decoder.}
\label{fig:AdvLN}
\end{figure*}

Figures~\ref{fig:NaiveLN} and \ref{fig:AdvLN} show sample results of logit noising for a benign audio wave and an adversarial example.
The audio adversarial example shows that many tokens are inverted compared with the benign audio wave, which is relatively robust to logit noising.
To effectively detect audio adversarial examples, we developed a strategy to select appropriate noise and a detailed detection method. 
\\

\PP{Noise Selection.} First, we calculate the probability that the largest logit value and the $k^{th}$ largest logit value are inverted as a result of random noise added to each logit. Let $L_{1}$ and $L_{k}$ be the values of the largest and the $k^{th}$ largest logit, respectively. Then, the probability that the logit values are inverted owing to the added noise can be represented as follows:  \\

$$
\begin{aligned}
P^{1,k}_{inv}(w) &= \sum_{m=1}^{n} (P_w(L_{1}-L_{k} = c_{m}) 
\\& \hspace{8mm} \cdot P(c_{m} < (\epsilon(L_{k}) - \epsilon(L_{1}))~|~L_{1}-L_{k} = c_{m}))\\
&=\sum_{m=1}^{n} P_w(L_{1}-L_{k} = c_{m}) \int_{-\infty}^{\infty} P_{\epsilon}(x) \int_{x+c_{m}}^{\infty}P_{\epsilon}(y)dydx  \\
\end{aligned}
$$
where $\epsilon()$, $P_{\epsilon}()$, $P_w()$, and $n$ represent the noise, probability distribution of noise, probability distribution of logit values for the given input wave type $w$, and the number of logits, respectively. 
Then we determine the total inversion probability as follows:
$$
P^{total}_{inv}(w) = (1- \prod_{k=2} (1- P^{1,k}_{inv}(w))) 
$$
Using the above equation, we can calculate the probability of inversion in various logit values as long as $P_{\epsilon}()$ and $P_w()$ are known.
We obtain $P_w$ by performing ASR inferences using sample waves (both adversarial and benign). 
For $P_{\epsilon}()$, we determine a proper noise through the following analysis.

Determining a proper distribution for noise that inverts the logits of adversarial examples more than that of benign waves is challenging.
We use Gaussian-distributed noise in this study, with the mean of the Gaussian distribution set at zero.
As the largest logit is close to the second and third largest logit in the adversarial examples (Figure~\ref{fig:5}), subtle noise may induce inversions of adversarial examples without inverting the benign waves.
Therefore, by varying the standard deviation ($std$) of the distribution, we are able to determine the best $std$ for distinguishing adversarial examples.
The problem is formulated as follows:
$$
\begin{aligned}
Find\hspace{1mm} std(P_{\epsilon})\hspace{1mm} s.t.\hspace{1mm} P^{total}_{inv}(adv.) \hspace{1mm} \gg \hspace{1mm} P^{total}_{inv}(benign) \\
\end{aligned}
$$
where $std(P_{\epsilon})$ represents the $std$ of the noise distribution $P_{\epsilon}$.
The analysis of the experiments is explained in Subsection~\ref{sec:eval_noiseparameter}.

After determining the noise distribution, we apply the noise to logits and feed the results into the DE to produce a transcription result.
We then compare the results with and without the logit noising.
Let us denote $ASR(x) = DE(AM(x))$, then our detection method can be modeled as follows:

$$
  \begin{cases}
    ~~DE(AM(x)+P_{\epsilon}) \simeq DE(AM(x)) &\mbox{if } x \ne x^{adv}\\ 
    ~~DE(AM(x^{adv})+P_{\epsilon}) \ne DE(AM(x^{adv})) &\mbox{otherwise}\\
  \end{cases}
$$
\\

\subsection{Detecting Adversarial Examples}
\label{sec:dae}
To determine whether a given input is an adversarial example or not, we measure the errors of the transcription result arising from the logit noising.
We compute character error rate (CER) between the transcription results with and without logit noising.
The CER is a metric that compares how different the character of generated sentence is from those of the original (ground-truth) sentence based on the Levenshtein distance~\cite{levenshtein1966binary}.
The CER between two sentences is calculated using the following equation:
$$
CER(S_1, S_2) = \frac{I + S + D}{N}
$$
where $S_1$ and $S_2$ represent the original sentence characters and the generated sentence characters, respectively. In addition, $I$ is the number of insertion, $S$ is the number of substitution, $D$ is the number of deletion, and $N$ denotes the number of original sentence characters. 
We determine the optimal CER threshold via experiment.

\PP{Multiple instances of logit noising.}
As a noise sampled from a Gaussian distribution has randomness, the average result of multiple instances of logit noising can further increase the detection accuracy.
First, we create multiple instances of logit noising by sampling multiple noises and adding them to logits.
We then generate transcription results for each logit noising instance.
We compute CERs between the transcription results of logit noising instances and the original logit.
If the average CER falls below the threshold, the proposed system determines that the input wave is an adversarial example.
In this paper, we empirically determine the number of logits noising instances to decide whether the given input is adversarial or not, and it is set to 4.

\section{Experimental Evaluation}

\subsection{Experimental Setup}

We evaluate the effectiveness of the proposed method by applying it to \textit{\textbf{DeepSpeech}}\footnote{https://github.com/mozilla/DeepSpeech} ~\cite{hannun2014deep}, which is Mozilla's open source ASR system.
We implement the proposed method using TensorFlow~\cite{tensorflow} and conduct all experiments on a workstation with one NVIDIA TITAN V GPU.

We use three over-line attacks~\cite{C_W,weighted_aaai20,blackboxSPW19} and two over-air attacks~\cite{hiromu,tao2020Metamorph} to evaluate the robustness of the proposed detection system.
For the over-line attacks, we feed adversarial examples in a digital format (.wav file) directly to DeepSpeech (v0.1.0).
We then compute the CER caused by logit noising.
For the over-air attacks, we play the adversarial examples through the speaker (DELL N889) and record the sound in a small room using a receiver that covers 0.5 m (Samsung Galaxy S8).
Note that, we also play benign waves through the speaker and record as the same with adversarial examples for a fair comparison.
Then, we compute CER and measure accuracy, false positive rate (FPR), and false negative rate (FNR) to evaluate the effectiveness of detection method. 
FPR is the rate at which a method mistakes benign audio waves for false adversarial examples and FNR is the rate at which a method misses false benign audio waves (actually adversarial).

We generate adversarial examples using the authors' source code if the code is available (C\&W attack~\footnote{https://github.com/carlini/audio\_adversarial\_examples}, Taori attack~\footnote{https://github.com/rtaori/Black-Box-Audio}, and Hiromu attack\footnote{https://github.com/hiromu/robust\_audio\_ae}); otherwise, we use the publicly available adversarial examples generated by the authors (weight-sampling attack\footnote{https://sites.google.com/view/audio-adversarial-examples} and metamorph attack\footnote{https://acoustic-metamorph-system.github.io/}).
We also reproduce three state-of-the-art detection methods~\cite{Kwon19ccsw,yang2018characterizing,MM20detection} for performance comparison.\\

\noindent\textbf{Dataset.}
We use \textit{LibriSpeech}\footnote{https://www.openslr.org/12} dataset~\cite{LibriSpeech}.
The \textit{LibriSpeech} dataset consists of a large corpus of 16kHz English audio data from the LibriVox project.
We use a \textit{test-clean} dataset of LibriSpeech for generating audio adversarial examples~\cite{C_W,hiromu,blackboxSPW19}.
We split the dataset into a set for determining the parameters and a test dataset.

The first set consists of 50 \textit{LibriSpeech} data, and we generate adversarial examples of C\&W and Hiromu attacks using the dataset.
We then search for optimal parameters ($std(P_{\epsilon})$ and CER threshold) of the method.
Detailed experiments for the parameter selection are explained in Section~\ref{sec:eval_noiseparameter}.

The second set (test set) consists of 500 long data points and 500 short data points.
We use 500 long data points to generate adversarial examples of C\&W attack and Hiromu attack.
These attacks are capable of generating adversarial examples with long audio data, which is beneficial for long target sentences.
On the other hand, the Taori attack is not suitable for long audio samples because of its high computing overhead~\cite{blackboxSPW19}.
For the Taori attack, we use 500 short data points to generate adversarial examples.
The average lengths of the long and short data are 4.6s and 2.2s, respectively.
We select target sentences based on the time length of the audio data.
Target sentences for long and short data consist of three to five words and two words, respectively~(See Table~\ref{tab:ATS}).
The details of the datasets used in the experiments are summarized in Table~\ref{tab:my_label}.

\begin{table}[t]
    \centering
    \caption{Target sentences used to generate adversarial examples}
    \label{tab:ATS}
    \begin{tabular}{c|l}
    % \hline
    % \multicolumn{2}{c}{\textbf{Adversarial Target Sentences}}                 \\ 
    \hline
    \multirow{5}{*}{\textbf{Long target sentences}} & open the door                      \\
                           & airplane mode on                   \\
                           & turn off the light                 \\
                           & this is adversarial example     \\
                           & clear all appointments on calendar \\ \hline
    \multirow{2}{*}{\textbf{Short target sentences}}      & ok google                          \\
           & hello world                        \\ \hline
    \end{tabular}
\end{table}
\begin{table}[h]
    \centering
    %\begin{threeparttable}
    \caption{Evaluation dataset composition}
    \label{tab:my_label}
    \begin{tabular}{c|c|c|c}
    \hline
    \multicolumn{2}{c|}{\textbf{Attack Method}} & \textbf{benign} & \textbf{attack} \\
    \cline{1-4}
    \multirow{3}{*}{Over-line}     & C\&W &500 (long) &500\\
    \cline{2-4}
                         &weighted-sampling* & 6& 11\\
    \cline{2-4}
                        &Taori & 500 (short) & 500\\
    \cline{1-4}
    
    \multirow{2}{*}{Over-air} & Hiromu & 500 (long) & 500 \\
    \cline{2-4}
                        &Metamorph* & 4 & 16 \\
    \hline
    \multicolumn{4}{l}{\small * We use publicly available adversarial examples.} \\
    \end{tabular}

\end{table} 

\begin{figure}[t]
\begin{center}
  \includegraphics[width=.95\linewidth]{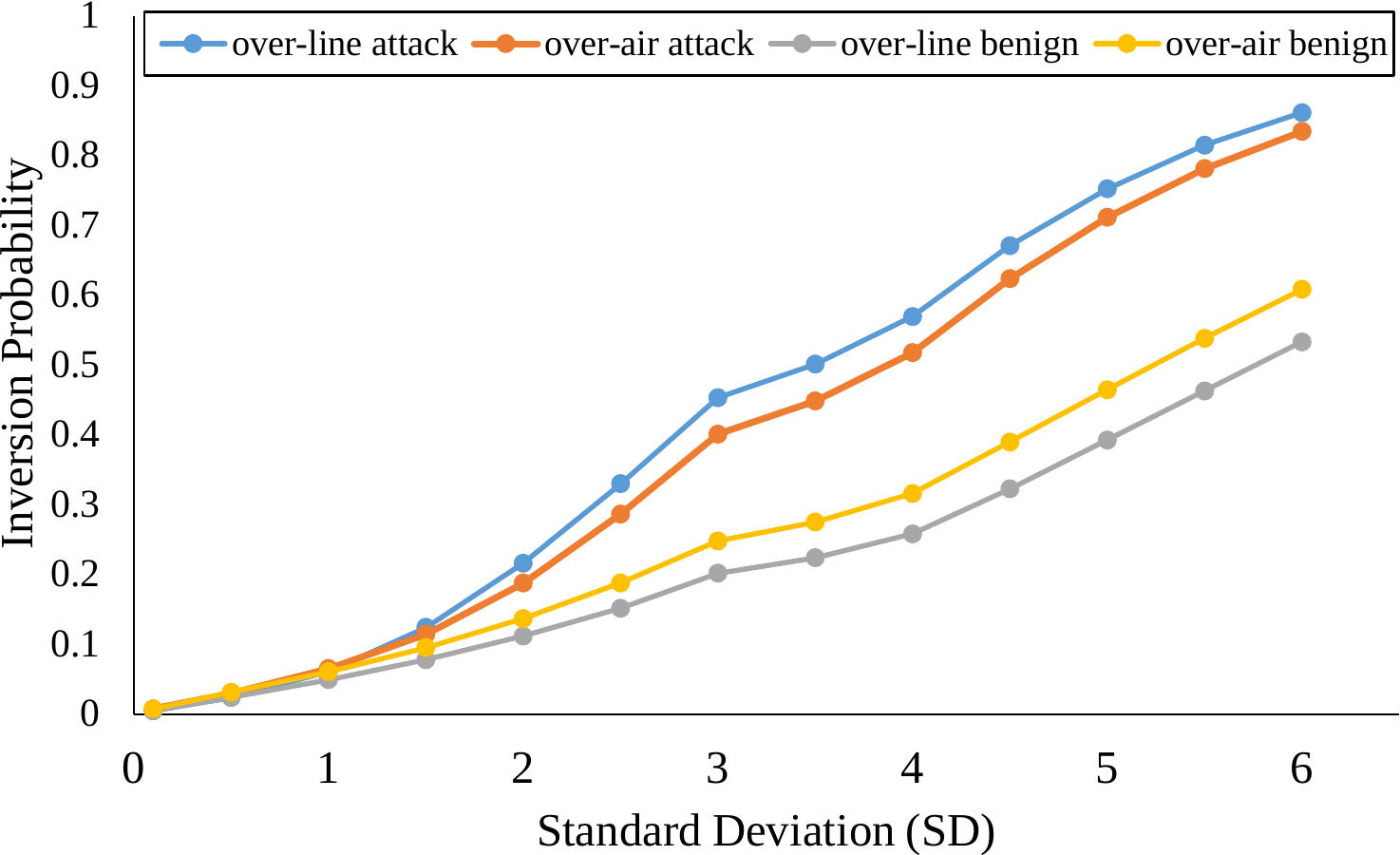}
\end{center}
\caption{$P_{inv}$ for different input wave types with varying $std$. We use 50 LibirSpeech dataset and 50 C\&W attack samples for over-line, and use 50 over-air LibriSpeech dataset and 50 Hiromu attack samples.}
\label{fig:NoiseSD}
\end{figure}

\begin{figure}%
    \centering
    \subfloat[The CER score in over-line\label{fig:overline_CER}]{{\includegraphics[width=8cm]{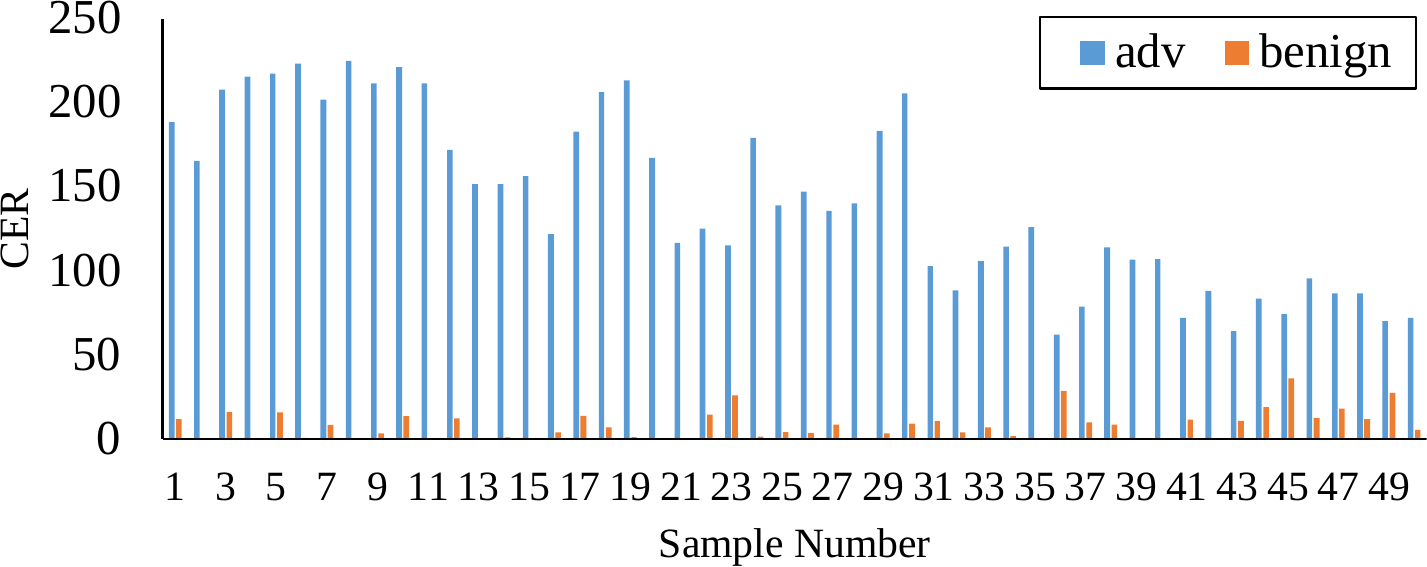} }}%
    \qquad
    \subfloat[The CER score in over-air\label{fig:overair_CER}]{{\includegraphics[width=8cm]{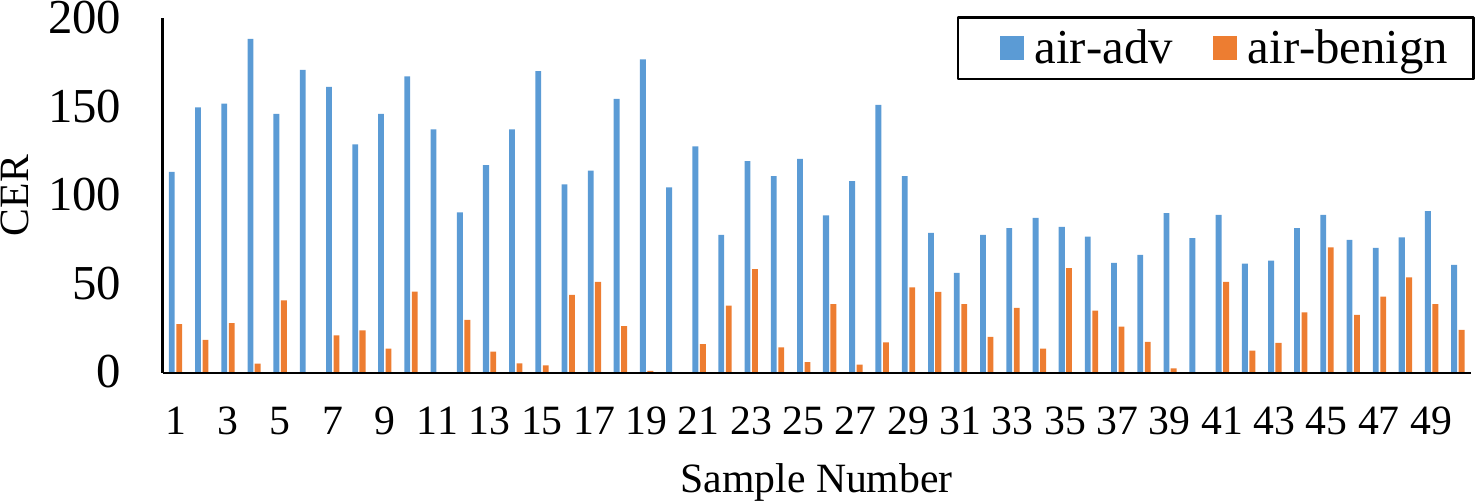} }}%
    \caption{The CER evaluation in over-line \& over-air situation. We use 50 LibirSpeech dataset and 50 C\&W attack samples for over-line situation, and use 50 over-air LibriSpeech dataset and 50 Hiromu attack samples.}%
\end{figure}

\subsection{Noise Parameter Selection}
\label{sec:eval_noiseparameter}
We set up $std$ of Gaussian noise and CER threshold that can effectively detect audio adversarial examples while having a low FPR. 
Low FPR is particularly important because an ASR system has to recognize benign audio as benign for maintaining its usability~\cite{abdullah2020sok}.
We determine the parameters as follows.

First, we calculate the inversion probability ($P_{inv}$) of four types of waves, varying the $std$ values of Gaussian noise distribution (Figure~\ref{fig:NoiseSD}). 
For this experiment, we use 50 benign waves, over-air benign waves, C\&W (over-line) attack examples, and Hiromu attack examples (over-air).
As can be observed from Figure~\ref{fig:NoiseSD}, the $P_{inv}$ difference between the audio adversarial examples and benign samples is noticeably different from when the $std$ is 3 (audio adversarial examples (0.4 and 0.45) and benign samples (0.02 and 0.24)), and the difference decreases from when the $std$ is 5 (audio adversarial examples (0.71 and 0.75) and benign samples (0.39 and 0.46)).
We choose $std$ candidates (2-5) whose $P_{inv}$ is low for benign waves and high for adversarial examples.
Subsequently, We search for the optimal CER threshold for each $std$ candidates.
For example, Figure~\ref{fig:overline_CER} shows the results of over-line adversarial examples when $std$ is 3, and Figure~\ref{fig:overair_CER} shows the results of over-air adversarial examples. 
We use a grid search to find the optimal CER and obtain the pairs of CER threshold and $std$.
We show FPR and FNR against the over-line and over-air adversarial examples (Table~\ref{tab:NoiseSt}). 
From the table, FPR is proportional to $std$, and FNR is inversely proportional to $std$.
It is necessary to set an appropriate pair of $std$ and CER threshold, and we decide to use 3 for $std$ and 60 for CER threshold.

\color{black}
\begin{table}[h]
    \centering
    \caption{FPR and FNR for different pairs of $std$ and CER threshold against adversarial example samples}
    \label{tab:NoiseSt}
    \begin{tabular}{c|c|c|c|c}
    \hline
    \multirow{2}{*}{\textbf{Eval.}} & \multicolumn{4}{c}{($std$, CER threshold)}\\
    \cline{2-5}
        & (2, 30) & (3, 60) & (4, 70) & (5, 110) \\
    \cline{1-5}
    Acc. & 175 (87.5\%) & 196 (98\%) & 184 (92\%) & 185 (92.5\%) \\
    \cline{1-5}
    FPR & 4 (4\%) & 3 (3\%) & 16 (16\%) & 14 (14\%) \\
    \cline{1-5}
    FNR & 21 (21\%) & 1 (1\%) & 0 (0\%) & 1 (1\%) \\
    \hline
    \end{tabular}
\end{table}

As we mentioned Section~\ref{sec:dae}, we consider multiple instances of logit noising for a given input. 
We show the effect of multiple instance of logit noising in Table~\ref{tab:threshold}. 
We first calculate the CER of each instance and averaged them to decide whether the input is benign or adversarial example. 
The table shows FPR and FNR against the 500 over-air adversarial examples and their corresponding benign examples.
 
It can be seen that as the number of logit noising instances increases, the FNR and FPR decrease.
As large number of logit noising instances induces additional overhead, we decide to use 4 instances of logit noising for a transcription.
Multiple instances of logit noising make stable result, thus contributing to making a robust system.

\begin{table}[h]
    \centering
    \caption{FPR and FNR for different number of noised instances}
    \label{tab:threshold}
    \begin{tabular}{c|c|c|c|c}
    \hline
    \multirow{2}{*}{\textbf{Eval.}} & \multicolumn{4}{c}{number of noised instance ($std$=3, CER threshold=60)}\\
    \cline{2-5}
        & 1 & 2 & 4 & 8 \\
    \cline{1-5}
    Acc. & 987 (98.7\%) & 988 (98.7\%) & 990 (99\%) & 996 (99.6\%) \\
    \cline{1-5}
    FPR & 3 (0.6\%) & 3 (0.6\%) & 3 (0.6\%) & 2 (0.4\%) \\
    \cline{1-5}
    FNR & 10 (2\%) & 9 (1.8\%) & 7 (1.4\%) & 2 (0.4\%) \\
    \hline
    \end{tabular}
    
\end{table}

\subsection{Detection Accuracy}

Table~\ref{tab:attack_evaluation} shows the performance of existing methods and the proposed method against various attacks.
In the table, R.N.P.~\cite{Kwon19ccsw} represents the random noise padding method which detects attacks by modifying input waves using a low pass filter. 
TD1 and TD2~\cite{yang2018characterizing} represent the detection method that split input audio into 5:5 or 4:1 to find temporal dependencies between the split components.
Reverb.~\cite{MM20detection} represents the detection method that adds reverberation noise to the input waves.
Reverb. uses 4 different types of reverberation randomly for detection.
For ours, we set $std(P_{\epsilon})$ to 3 and use four logit noising instances as above-mentioned.
Each detection method uses different metrics to decide whether the input is benign or adversarial.
Reverb. uses word error rate (WER), whereas TD1 and TD2 use character error rate (CER). 
We use CER instead of WER because the proposed method modifies logits, and that will directly affect each character unit.

\begin{table*}
\centering
\caption{Evaluation of various detection methods for different attacks.}
\label{tab:attack_evaluation}
\begin{tabular}{|c|c|c|c|c|c|c|c|}
\hline
\multirow{2}{*}{Type} &\multirow{2}{*}{\textbf{Attack (Dataset)}} & \multirow{2}{*}{\textbf{Performance}} & \multicolumn{5}{c|}{\textbf{Detection method}}\\
\cline{4-8}
& & & R. N. P.~\cite{Kwon19ccsw} & TD1 (5:5)~\cite{yang2018characterizing} &  TD2 (4:1)~\cite{yang2018characterizing} & Reverb. (n=4)~\cite{MM20detection} & \textbf{ours ($std$=3)} \\
\hline
\multirow{9}{*}{Over-line} & & 
                             Accuracy & \textbf{1000 (100\%)} & 923 (92.3\%) & 958 (95.8\%) & \textbf{1000 (100\%)} & \textbf{1000 (100\%)} \\
                         \cline{3-8}
                         &  C\&W ($D_{long}$) ~\cite{C_W} & FPR & 0 (0\%) & 6 (1.2\%) & 3 (0.6\%) & 0 (0\%) & 0 (0\%)  \\
                         \cline{3-8}
                         & & FNR & 0 (0\%) & 71 (14.2\%) & 39 (7.8\%)  & 0 (0\%)   & 0 (0\%) \\
                         \cline{2-8}
                         &  & 
                             Accuracy & \textbf{17 (100\%)} & \textbf{17 (100\%)} & 15 (88.2\%) & \textbf{17 (100\%)} & \textbf{17 (100\%)} \\
                         \cline{3-8} 
                         & W-S (-) \cite{weighted_aaai20} & FPR & 0 (0\%) & 0 (0\%) & 0 (0\%) & 0 (0\%)& 0 (0\%) \\
                         \cline{3-8}
                         & & FNR & 0 (0\%) & 0 (0\%) & 2 (18.2\%) & 0 (0\%)  & 0 (0\%)  \\
                         \cline{2-8}
                         & & 
                             Accuracy & 998 (99.8\%) & 847 (84.7\%) & 889 (88.9\%) & \textbf{1000 (100\%)} & \textbf{1000 (100\%)} \\
                         \cline{3-8} 
                         & Taori ($D_{short}$) \cite{blackboxSPW19} & FPR & 0 (0\%) & 115 (23.0\%) & 100 (20.0\%) & 0 (0\%) & 0 (0\%)  \\
                         \cline{3-8}
                         & & FNR & 2 (0.4\%)  & 38 (7.6\%) & 11 (2.2\%) & 0 (0\%)  & 0 (0\%) \\
                         \cline{1-8}
\multirow{6}{*}{Over-air} & &
                            Accuracy & 507 (50.7\%) & 679 (67.9\%) & 691 (69.1\%) & 937 (93.7\%) & \textbf{990 (99.0\%)} \\
                            \cline{3-8}  
                         & Hiromu* ($D_{long}$) \cite{hiromu} & FPR & 0 (0\%) & 27 (5.4\%) & 3 (0.6\%) & 10 (2.0\%) & 3 (0.6\%)\\
                            \cline{3-8}
                         & & FNR & 493 (98.6\%) & 294 (58.8\%) & 306 (61.2\%) & 53 (10.6\%) & 7 (1.4\%)\\
                            \cline{2-8}
                         & & Accuracy & 4 (20\%) & 9 (45.0\%) & 11 (55\%) & 13 (65\%) & \textbf{19 (95\%)} \\
                         \cline{3-8} 
                         & Metamorph* (-)~\cite{tao2020Metamorph} & FPR & 0 (0\%) & 0 (0\%) & 0 (0\%) & 0 (0\%) & 0 (0\%)  \\
                         \cline{3-8}
                         & & FNR & 16 (100\%) & 11 (68.75\%) & 9 (56.25\%)  & 7 (43.75\%)  & 1 (6.25\%) \\
                        \hline
\multicolumn{4}{l}{\small * means dataset recorded over-air.} \\
\end{tabular}
\end{table*}

From Table~\ref{tab:attack_evaluation}, it can be observed that most detection methods exhibit high detection accuracy for over-line attacks. 
For C\&W and W-S attacks, the FNRs and FPRs of all detection methods are 0\% (Acc.: 100\%) except for TD1 and TD2. 
Although the TD1 and TD2 methods cannot detect all the over-line attacks, TD1 and TD2 methods show high accuracy (TD1: 92.3 \% \& TD2: 95.8 \%). 
Against the Taori attack, only the Reverb. and the proposed method achieve 0\% FPR and 0\% FNR.
Although other methods suffer from high false positives and false negatives, these methods achieve comparable accuracy.
These results imply that all the detection methods effectively detect adversarial examples of over-line attacks.

The most noticeable point is the result of over-air attacks.
R.N.P., TD1, and TD2 have high FNR (from 98.6\% to 58.8\% against the Hiromu attack and from 100\% to 56.25\% against the Metamorph attack).
The results indicate that these detection methods have limited performance at detecting over-air adversarial examples, although they recognize benign samples well.

In contrast to existing methods, the proposed method exhibit low FNR and FPR for all over-air attacks.
Furthermore, it shows a low FPR (0.6\% against the Hiromu attack and 6.25\% against the Metamorph attack).
These results demonstrate that our detection method is more robust than the other methods for over-air adversarial examples.

\section{Discussion}

Our experiments show that the proposed method, which adds noise to the logit, can detect audio adversarial examples effectively without additional components.\\

\PP{Logit Distribution Difference.}
The preceding analysis of our experiment shows that the logit-gap distribution of benign audio waves is different from that of audio adversarial examples.
The logit-gap of audio adversarial examples is densely distributed than that of benign waves.
We conjecture that the largest logit is for the target phrase and other logits are for the original wave when an adversarial example is successful.
As the adversarial example have to resemble human-audible original wave, the logit-gap of the audio adversarial examples is small.
This phenomenon makes the proposed method detect audio adversarial examples effectively.\\

\PP{Noise Distribution.} In this paper, we only use Gaussian noise distribution. We think that there are some noise distributions that may and may not work depending on the distribution of the gap value. We consider that it may need a tailored noise distribution that will separate benign samples and adversarial examples.\\

\PP{Type of Source Wave.} We have shown results experimented with a human voice in LibriSpeech dataset. As attacks with various input wave types have been proposed~\cite{yuan2018commandersong,li2019music}, we have also conducted experiments on song wave type. In this experiment, we only focus on false negative rate (FNR). This is because the song may not contain any transcription, and the target model (DeepSpeech v0.1.0) does not recognize song with a lyric well. The logit noising method is able to detect song based audio adversarial examples well (FNR: 98.11 \%).\\

\PP{ASR Platform.}
We have conveyed our study on DeepSpeech platform. We have not considered attacks built on other platforms like Kaldi~\cite{povey2011kaldi} \& Lingvo~\cite{lingvo}.
However, other platforms also adopt MFCC as acoustic model's input, and the proposed method only uses acoustic model result (logits). Therefore, we believe that our method also works on them. \\

\PP{Future work.}
One thing we carefully consider is whether our method can be used to recover original sentences from adversarial examples. We have experimented with different distribution of Gaussian noise. There are some cases that have recovered original transcription results when the perturbed signal in adversarial examples is subtle. However, when there is a huge perturbation to make adversarial examples, the sound is a little noisy and it is unable to recover original transcription from logit noising. We think it needs further study to recover the original transcription using logit noising.

Adaptive attacks can be considered for future work. Although our  method has a great effect on various types of attack, we do not consider adaptive attackers who know the logit noising detection method. However, we predict that it is difficult to make robust logit to noise for adversarial examples, not like benign samples, through adaptive attacks because considering both generating target transcription from continuous vector frames and making the largest logit value and another logit values require high computational cost. 
\section{Conclusion}

In this paper, we introduce a novel and simple audio adversarial example detection method applicable to speech recognition systems. 
We discover that audio adversarial examples are distinguishable by adding noise to logits before feeding them into the ASR decoder. 
To evaluate our proposed system, we experiment with an open-source speech recognition system, DeepSpeech. 
We evaluate various attacks, including three over-line attacks and two over-air attacks. 
We demonstrate that the accuracy of our method is higher than that of existing state-of the art methods. 
Most importantly, our method is effective at detecting over-air attacks. 
The proposed method is easy to implement and does not require model retraining. 
We expect our results to provide another avenue to detect audio adversarial attacks.

\begin{acks}
We would like to thank the anonymous reviewers for their invaluable comments and suggestions.
This research was supported by IITP-2018-0-01392 and IITP-2018-0-01441 through the Institute of Information and Communication Technology Planning and Evaluation (IITP) funded by the Ministry of Science and ICT.
\end{acks}

\bibliographystyle{ACM-Reference-Format}
\bibliography{reference.bib}


\begin{thebibliography}{36}

%%% the Web address.

\ifx \showCODEN    \undefined \def \showCODEN     #1{\unskip}     \fi
\ifx \showDOI      \undefined \def \showDOI       #1{#1}\fi
\ifx \showISBNx    \undefined \def \showISBNx     #1{\unskip}     \fi
\ifx \showISBNxiii \undefined \def \showISBNxiii  #1{\unskip}     \fi
\ifx \showISSN     \undefined \def \showISSN      #1{\unskip}     \fi
\ifx \showLCCN     \undefined \def \showLCCN      #1{\unskip}     \fi
\ifx \shownote     \undefined \def \shownote      #1{#1}          \fi
\ifx \showarticletitle \undefined \def \showarticletitle #1{#1}   \fi
\ifx \showURL      \undefined \def \showURL       {\relax}        \fi
% invisible to TeX
\providecommand\bibfield[2]{#2}
\providecommand\bibinfo[2]{#2}
\providecommand\natexlab[1]{#1}
\providecommand\showeprint[2][]{arXiv:#2}

\bibitem[\protect\citeauthoryear{Abadi et~al\mbox{.}}{Abadi
  et~al\mbox{.}}{2015}]%
        {tensorflow}
\bibfield{author}{\bibinfo{person}{Mart\'{\i}n Abadi} {et~al\mbox{.}}}
  \bibinfo{year}{2015}\natexlab{}.
\newblock \bibinfo{title}{{TensorFlow}: Large-Scale Machine Learning on
  Heterogeneous Systems}.
\newblock
\newblock
\urldef\tempurl%
\url{https://www.tensorflow.org/}
\showURL{%
\tempurl}
\newblock
\shownote{Software available from tensorflow.org}.


\bibitem[\protect\citeauthoryear{Abdullah, Rahman, Garcia, Warren, Yadav,
  Shrimpton, and Traynor}{Abdullah et~al\mbox{.}}{2021}]%
        {advertraining@sp21}
\bibfield{author}{\bibinfo{person}{H. Abdullah}, \bibinfo{person}{M. Rahman},
  \bibinfo{person}{W. Garcia}, \bibinfo{person}{K. Warren},
  \bibinfo{person}{A.~Swarnim Yadav}, \bibinfo{person}{T. Shrimpton}, {and}
  \bibinfo{person}{P. Traynor}.} \bibinfo{year}{2021}\natexlab{}.
\newblock \showarticletitle{Hear "No Evil", See "Kenansville"*: Efficient and
  Transferable Black-Box Attacks on Speech Recognition and Voice Identification
  Systems}. In \bibinfo{booktitle}{\emph{2021 2021 IEEE Symposium on Security
  and Privacy (SP)}}. \bibinfo{publisher}{IEEE Computer Society},
  \bibinfo{address}{Los Alamitos, CA, USA}, \bibinfo{pages}{142--159}.
\newblock
\showISSN{2375-1207}
\urldef\tempurl%
\url{https://doi.org/10.1109/SP40001.2021.00009}
\showDOI{\tempurl}


\bibitem[\protect\citeauthoryear{Abdullah, Warren, Bindschaedler, Papernot, and
  Traynor}{Abdullah et~al\mbox{.}}{2020}]%
        {abdullah2020sok}
\bibfield{author}{\bibinfo{person}{Hadi Abdullah}, \bibinfo{person}{Kevin
  Warren}, \bibinfo{person}{Vincent Bindschaedler}, \bibinfo{person}{Nicolas
  Papernot}, {and} \bibinfo{person}{Patrick Traynor}.}
  \bibinfo{year}{2020}\natexlab{}.
\newblock \bibinfo{title}{SoK: The Faults in our ASRs: An Overview of Attacks
  against Automatic Speech Recognition and Speaker Identification Systems}.
\newblock
\newblock
\showeprint[arxiv]{2007.06622}~[cs.CR]


\bibitem[\protect\citeauthoryear{Ahmed, Natarajan, and Rao}{Ahmed
  et~al\mbox{.}}{1974}]%
        {DCT}
\bibfield{author}{\bibinfo{person}{Nasir Ahmed}, \bibinfo{person}{T\_
  Natarajan}, {and} \bibinfo{person}{Kamisetty~R Rao}.}
  \bibinfo{year}{1974}\natexlab{}.
\newblock \showarticletitle{Discrete cosine transform}.
\newblock \bibinfo{journal}{\emph{IEEE transactions on Computers}}
  \bibinfo{volume}{100}, \bibinfo{number}{1} (\bibinfo{year}{1974}),
  \bibinfo{pages}{90--93}.
\newblock


\bibitem[\protect\citeauthoryear{{"Amazon"}}{{"Amazon"}}{[n.\,d.]}]%
        {Alexa}
\bibfield{author}{\bibinfo{person}{{"Amazon"}}.}
  \bibinfo{year}{[n.\,d.]}\natexlab{}.
\newblock \bibinfo{title}{{"Amazon Alexa"}}.
\newblock \bibinfo{howpublished}{\url{https://www.amazon.com}}.
\newblock


\bibitem[\protect\citeauthoryear{{"Apple"}}{{"Apple"}}{[n.\,d.]}]%
        {Siri}
\bibfield{author}{\bibinfo{person}{{"Apple"}}.}
  \bibinfo{year}{[n.\,d.]}\natexlab{}.
\newblock \bibinfo{title}{{"Apple Siri"}}.
\newblock \bibinfo{howpublished}{\url{https://www.apple.com/siri}}.
\newblock


\bibitem[\protect\citeauthoryear{Athalye, Engstrom, Ilyas, and Kwok}{Athalye
  et~al\mbox{.}}{2018}]%
        {EOT18ICML}
\bibfield{author}{\bibinfo{person}{Anish Athalye}, \bibinfo{person}{Logan
  Engstrom}, \bibinfo{person}{Andrew Ilyas}, {and} \bibinfo{person}{Kevin
  Kwok}.} \bibinfo{year}{2018}\natexlab{}.
\newblock \showarticletitle{Synthesizing Robust Adversarial Examples}. In
  \bibinfo{booktitle}{\emph{Proceedings of the 35th International Conference on
  Machine Learning}} \emph{(\bibinfo{series}{Proceedings of Machine Learning
  Research}, Vol.~\bibinfo{volume}{80})},
  \bibfield{editor}{\bibinfo{person}{Jennifer Dy} {and}
  \bibinfo{person}{Andreas Krause}} (Eds.). \bibinfo{publisher}{PMLR},
  \bibinfo{pages}{284--293}.
\newblock
\urldef\tempurl%
\url{http://proceedings.mlr.press/v80/athalye18b.html}
\showURL{%
\tempurl}


\bibitem[\protect\citeauthoryear{Carlini and Wagner}{Carlini and
  Wagner}{2018}]%
        {C_W}
\bibfield{author}{\bibinfo{person}{Nicholas Carlini} {and}
  \bibinfo{person}{David Wagner}.} \bibinfo{year}{2018}\natexlab{}.
\newblock \showarticletitle{Audio Adversarial Examples: Targeted Attacks on
  Speech-to-Text}. In \bibinfo{booktitle}{\emph{2018 IEEE Security and Privacy
  Workshops (SPW)}}. \bibinfo{pages}{1--7}.
\newblock
\urldef\tempurl%
\url{https://doi.org/10.1109/SPW.2018.00009}
\showDOI{\tempurl}


\bibitem[\protect\citeauthoryear{Chen, Shangguan, Li, and Jamieson}{Chen
  et~al\mbox{.}}{2020a}]%
        {tao2020Metamorph}
\bibfield{author}{\bibinfo{person}{Tao Chen}, \bibinfo{person}{Longfei
  Shangguan}, \bibinfo{person}{Zhenjiang Li}, {and} \bibinfo{person}{Kyle
  Jamieson}.} \bibinfo{year}{2020}\natexlab{a}.
\newblock \showarticletitle{Metamorph: Injecting Inaudible Commands into
  Over-the-air Voice Controlled Systems}. In
  \bibinfo{booktitle}{\emph{Proceedings of NDSS}}.
\newblock


\bibitem[\protect\citeauthoryear{Chen, Yuan, Zhang, Zhao, Zhang, Chen, and
  Wang}{Chen et~al\mbox{.}}{2020b}]%
        {develsecurity@2020}
\bibfield{author}{\bibinfo{person}{Yuxuan Chen}, \bibinfo{person}{Xuejing
  Yuan}, \bibinfo{person}{Jiangshan Zhang}, \bibinfo{person}{Yue Zhao},
  \bibinfo{person}{Shengzhi Zhang}, \bibinfo{person}{Kai Chen}, {and}
  \bibinfo{person}{XiaoFeng Wang}.} \bibinfo{year}{2020}\natexlab{b}.
\newblock \showarticletitle{Devil{\textquoteright}s Whisper: A General Approach
  for Physical Adversarial Attacks against Commercial Black-box Speech
  Recognition Devices}. In \bibinfo{booktitle}{\emph{29th {USENIX} Security
  Symposium ({USENIX} Security 20)}}. \bibinfo{publisher}{{USENIX}
  Association}, \bibinfo{pages}{2667--2684}.
\newblock
\showISBNx{978-1-939133-17-5}
\urldef\tempurl%
\url{https://www.usenix.org/conference/usenixsecurity20/presentation/chen-yuxuan}
\showURL{%
\tempurl}


\bibitem[\protect\citeauthoryear{D\"{o}rr, Markert, M\"{u}ller, and
  B\"{o}ttinger}{D\"{o}rr et~al\mbox{.}}{2020}]%
        {adv_training3}
\bibfield{author}{\bibinfo{person}{Tom D\"{o}rr}, \bibinfo{person}{Karla
  Markert}, \bibinfo{person}{Nicolas~M. M\"{u}ller}, {and}
  \bibinfo{person}{Konstantin B\"{o}ttinger}.} \bibinfo{year}{2020}\natexlab{}.
\newblock \showarticletitle{Towards Resistant Audio Adversarial Examples}. In
  \bibinfo{booktitle}{\emph{Proceedings of the 1st ACM Workshop on Security and
  Privacy on Artificial Intelligence}} (Taipei, Taiwan)
  \emph{(\bibinfo{series}{SPAI '20})}. \bibinfo{publisher}{Association for
  Computing Machinery}, \bibinfo{address}{New York, NY, USA},
  \bibinfo{pages}{3–10}.
\newblock
\showISBNx{9781450376112}
\urldef\tempurl%
\url{https://doi.org/10.1145/3385003.3410921}
\showDOI{\tempurl}


\bibitem[\protect\citeauthoryear{Du, Pun, and Zhang}{Du et~al\mbox{.}}{2020}]%
        {MM20detection}
\bibfield{author}{\bibinfo{person}{Xia Du}, \bibinfo{person}{Chi-Man Pun},
  {and} \bibinfo{person}{Zheng Zhang}.} \bibinfo{year}{2020}\natexlab{}.
\newblock \showarticletitle{A Unified Framework for Detecting Audio Adversarial
  Examples}. In \bibinfo{booktitle}{\emph{Proceedings of the 28th ACM
  International Conference on Multimedia}} (Seattle, WA, USA)
  \emph{(\bibinfo{series}{MM '20})}. \bibinfo{publisher}{Association for
  Computing Machinery}, \bibinfo{address}{New York, NY, USA},
  \bibinfo{pages}{3986–3994}.
\newblock
\showISBNx{9781450379885}
\urldef\tempurl%
\url{https://doi.org/10.1145/3394171.3413603}
\showDOI{\tempurl}


\bibitem[\protect\citeauthoryear{{"Google"}}{{"Google"}}{[n.\,d.]}]%
        {Assistant}
\bibfield{author}{\bibinfo{person}{{"Google"}}.}
  \bibinfo{year}{[n.\,d.]}\natexlab{}.
\newblock \bibinfo{title}{{"Google Assistant"}}.
\newblock \bibinfo{howpublished}{\url{https://assistant.google.com}}.
\newblock


\bibitem[\protect\citeauthoryear{Graves, Fern\'{a}ndez, Gomez, and
  Schmidhuber}{Graves et~al\mbox{.}}{2006}]%
        {CTC}
\bibfield{author}{\bibinfo{person}{Alex Graves}, \bibinfo{person}{Santiago
  Fern\'{a}ndez}, \bibinfo{person}{Faustino Gomez}, {and}
  \bibinfo{person}{J\"{u}rgen Schmidhuber}.} \bibinfo{year}{2006}\natexlab{}.
\newblock \showarticletitle{Connectionist Temporal Classification: Labelling
  Unsegmented Sequence Data with Recurrent Neural Networks}. In
  \bibinfo{booktitle}{\emph{Proceedings of the 23rd International Conference on
  Machine Learning}} (Pittsburgh, Pennsylvania, USA)
  \emph{(\bibinfo{series}{ICML '06})}. \bibinfo{publisher}{Association for
  Computing Machinery}, \bibinfo{address}{New York, NY, USA},
  \bibinfo{pages}{369–376}.
\newblock
\showISBNx{1595933832}
\urldef\tempurl%
\url{https://doi.org/10.1145/1143844.1143891}
\showDOI{\tempurl}


\bibitem[\protect\citeauthoryear{Hannun, Case, Casper, Catanzaro, Diamos,
  Elsen, Prenger, Satheesh, Sengupta, Coates, and Ng}{Hannun
  et~al\mbox{.}}{2014}]%
        {hannun2014deep}
\bibfield{author}{\bibinfo{person}{Awni Hannun}, \bibinfo{person}{Carl Case},
  \bibinfo{person}{Jared Casper}, \bibinfo{person}{Bryan Catanzaro},
  \bibinfo{person}{Greg Diamos}, \bibinfo{person}{Erich Elsen},
  \bibinfo{person}{Ryan Prenger}, \bibinfo{person}{Sanjeev Satheesh},
  \bibinfo{person}{Shubho Sengupta}, \bibinfo{person}{Adam Coates}, {and}
  \bibinfo{person}{Andrew~Y. Ng}.} \bibinfo{year}{2014}\natexlab{}.
\newblock \bibinfo{title}{Deep Speech: Scaling up end-to-end speech
  recognition}.
\newblock
\newblock
\showeprint[arxiv]{1412.5567}~[cs.CL]


\bibitem[\protect\citeauthoryear{Kwon, Yoon, and Park}{Kwon
  et~al\mbox{.}}{2019}]%
        {Kwon19ccsw}
\bibfield{author}{\bibinfo{person}{Hyun Kwon}, \bibinfo{person}{Hyunsoo Yoon},
  {and} \bibinfo{person}{Ki-Woong Park}.} \bibinfo{year}{2019}\natexlab{}.
\newblock \showarticletitle{POSTER: Detecting Audio Adversarial Example through
  Audio Modification}. In \bibinfo{booktitle}{\emph{Proceedings of the 2019 ACM
  SIGSAC Conference on Computer and Communications Security}} (London, United
  Kingdom) \emph{(\bibinfo{series}{CCS '19})}. \bibinfo{publisher}{Association
  for Computing Machinery}, \bibinfo{address}{New York, NY, USA},
  \bibinfo{pages}{2521–2523}.
\newblock
\showISBNx{9781450367479}
\urldef\tempurl%
\url{https://doi.org/10.1145/3319535.3363246}
\showDOI{\tempurl}


\bibitem[\protect\citeauthoryear{Levenshtein}{Levenshtein}{1966}]%
        {levenshtein1966binary}
\bibfield{author}{\bibinfo{person}{Vladimir~I Levenshtein}.}
  \bibinfo{year}{1966}\natexlab{}.
\newblock \showarticletitle{Binary codes capable of correcting deletions,
  insertions, and reversals}. In \bibinfo{booktitle}{\emph{Soviet physics
  doklady}}, Vol.~\bibinfo{volume}{10}. Soviet Union,
  \bibinfo{pages}{707--710}.
\newblock


\bibitem[\protect\citeauthoryear{Li, Qu, Li, Szurley, Kolter, and Metze}{Li
  et~al\mbox{.}}{2019}]%
        {li2019music}
\bibfield{author}{\bibinfo{person}{Juncheng~B Li}, \bibinfo{person}{Shuhui Qu},
  \bibinfo{person}{Xinjian Li}, \bibinfo{person}{Joseph Szurley},
  \bibinfo{person}{J~Zico Kolter}, {and} \bibinfo{person}{Florian Metze}.}
  \bibinfo{year}{2019}\natexlab{}.
\newblock \showarticletitle{Adversarial music: Real world audio adversary
  against wake-word detection system}.
\newblock \bibinfo{journal}{\emph{arXiv preprint arXiv:1911.00126}}
  (\bibinfo{year}{2019}).
\newblock


\bibitem[\protect\citeauthoryear{Li, Jiang, Wu, Hsieh, and Stolcke}{Li
  et~al\mbox{.}}{2020a}]%
        {adv_training1}
\bibfield{author}{\bibinfo{person}{Ruirui Li}, \bibinfo{person}{Jyun{-}Yu
  Jiang}, \bibinfo{person}{Xian Wu}, \bibinfo{person}{Chu{-}Cheng Hsieh}, {and}
  \bibinfo{person}{Andreas Stolcke}.} \bibinfo{year}{2020}\natexlab{a}.
\newblock \showarticletitle{Speaker Identification for Household Scenarios with
  Self-Attention and Adversarial Training}. In
  \bibinfo{booktitle}{\emph{Interspeech 2020, 21st Annual Conference of the
  International Speech Communication Association, Virtual Event, Shanghai,
  China, 25-29 October 2020}}, \bibfield{editor}{\bibinfo{person}{Helen Meng},
  \bibinfo{person}{Bo~Xu}, {and} \bibinfo{person}{Thomas~Fang Zheng}} (Eds.).
  \bibinfo{publisher}{{ISCA}}, \bibinfo{pages}{2272--2276}.
\newblock
\urldef\tempurl%
\url{https://doi.org/10.21437/Interspeech.2020-3025}
\showDOI{\tempurl}


\bibitem[\protect\citeauthoryear{Li, Wu, Liu, Chen, and Yuan}{Li
  et~al\mbox{.}}{2020b}]%
        {advpulse}
\bibfield{author}{\bibinfo{person}{Zhuohang Li}, \bibinfo{person}{Yi Wu},
  \bibinfo{person}{Jian Liu}, \bibinfo{person}{Yingying Chen}, {and}
  \bibinfo{person}{Bo Yuan}.} \bibinfo{year}{2020}\natexlab{b}.
\newblock \showarticletitle{AdvPulse: Universal, Synchronization-Free, and
  Targeted Audio Adversarial Attacks via Subsecond Perturbations}. In
  \bibinfo{booktitle}{\emph{Proceedings of the 2020 ACM SIGSAC Conference on
  Computer and Communications Security}} (Virtual Event, USA)
  \emph{(\bibinfo{series}{CCS '20})}. \bibinfo{publisher}{Association for
  Computing Machinery}, \bibinfo{address}{New York, NY, USA},
  \bibinfo{pages}{1121–1134}.
\newblock
\showISBNx{9781450370899}
\urldef\tempurl%
\url{https://doi.org/10.1145/3372297.3423348}
\showDOI{\tempurl}


\bibitem[\protect\citeauthoryear{Liu, Wan, Ding, Zhang, and Zhu}{Liu
  et~al\mbox{.}}{2020}]%
        {weighted_aaai20}
\bibfield{author}{\bibinfo{person}{Xiaolei Liu}, \bibinfo{person}{Kun Wan},
  \bibinfo{person}{Yufei Ding}, \bibinfo{person}{Xiaosong Zhang}, {and}
  \bibinfo{person}{Qingxin Zhu}.} \bibinfo{year}{2020}\natexlab{}.
\newblock \showarticletitle{Weighted-Sampling Audio Adversarial Example
  Attack}.
\newblock \bibinfo{journal}{\emph{Proceedings of the AAAI Conference on
  Artificial Intelligence}} \bibinfo{volume}{34}, \bibinfo{number}{04}
  (\bibinfo{date}{Apr.} \bibinfo{year}{2020}), \bibinfo{pages}{4908--4915}.
\newblock
\urldef\tempurl%
\url{https://doi.org/10.1609/aaai.v34i04.5928}
\showDOI{\tempurl}


\bibitem[\protect\citeauthoryear{Madry, Makelov, Schmidt, Tsipras, and
  Vladu}{Madry et~al\mbox{.}}{2017}]%
        {madry2017towards}
\bibfield{author}{\bibinfo{person}{Aleksander Madry},
  \bibinfo{person}{Aleksandar Makelov}, \bibinfo{person}{Ludwig Schmidt},
  \bibinfo{person}{Dimitris Tsipras}, {and} \bibinfo{person}{Adrian Vladu}.}
  \bibinfo{year}{2017}\natexlab{}.
\newblock \showarticletitle{Towards deep learning models resistant to
  adversarial attacks}.
\newblock \bibinfo{journal}{\emph{arXiv preprint arXiv:1706.06083}}
  (\bibinfo{year}{2017}).
\newblock


\bibitem[\protect\citeauthoryear{{"Microsoft"}}{{"Microsoft"}}{[n.\,d.]}]%
        {Cortana}
\bibfield{author}{\bibinfo{person}{{"Microsoft"}}.}
  \bibinfo{year}{[n.\,d.]}\natexlab{}.
\newblock \bibinfo{title}{{"Microsoft Cortana"}}.
\newblock
  \bibinfo{howpublished}{\url{https://www.microsoft.com/en-us/cortana}}.
\newblock


\bibitem[\protect\citeauthoryear{Muda, Begam, and Elamvazuthi}{Muda
  et~al\mbox{.}}{2010}]%
        {MFCC}
\bibfield{author}{\bibinfo{person}{Lindasalwa Muda}, \bibinfo{person}{Mumtaj
  Begam}, {and} \bibinfo{person}{Irraivan Elamvazuthi}.}
  \bibinfo{year}{2010}\natexlab{}.
\newblock \showarticletitle{Voice recognition algorithms using mel frequency
  cepstral coefficient (MFCC) and dynamic time warping (DTW) techniques}.
\newblock \bibinfo{journal}{\emph{arXiv preprint arXiv:1003.4083}}
  (\bibinfo{year}{2010}).
\newblock


\bibitem[\protect\citeauthoryear{Panayotov, Chen, Povey, and
  Khudanpur}{Panayotov et~al\mbox{.}}{2015}]%
        {LibriSpeech}
\bibfield{author}{\bibinfo{person}{Vassil Panayotov}, \bibinfo{person}{Guoguo
  Chen}, \bibinfo{person}{Daniel Povey}, {and} \bibinfo{person}{Sanjeev
  Khudanpur}.} \bibinfo{year}{2015}\natexlab{}.
\newblock \showarticletitle{Librispeech: An ASR corpus based on public domain
  audio books}. In \bibinfo{booktitle}{\emph{2015 IEEE International Conference
  on Acoustics, Speech and Signal Processing (ICASSP)}}.
  \bibinfo{pages}{5206--5210}.
\newblock
\urldef\tempurl%
\url{https://doi.org/10.1109/ICASSP.2015.7178964}
\showDOI{\tempurl}


\bibitem[\protect\citeauthoryear{Povey, Ghoshal, Boulianne, Burget, Glembek,
  Goel, Hannemann, Motlicek, Qian, Schwarz, et~al\mbox{.}}{Povey
  et~al\mbox{.}}{2011}]%
        {povey2011kaldi}
\bibfield{author}{\bibinfo{person}{Daniel Povey}, \bibinfo{person}{Arnab
  Ghoshal}, \bibinfo{person}{Gilles Boulianne}, \bibinfo{person}{Lukas Burget},
  \bibinfo{person}{Ondrej Glembek}, \bibinfo{person}{Nagendra Goel},
  \bibinfo{person}{Mirko Hannemann}, \bibinfo{person}{Petr Motlicek},
  \bibinfo{person}{Yanmin Qian}, \bibinfo{person}{Petr Schwarz},
  {et~al\mbox{.}}} \bibinfo{year}{2011}\natexlab{}.
\newblock \showarticletitle{The Kaldi speech recognition toolkit}. In
  \bibinfo{booktitle}{\emph{IEEE 2011 workshop on automatic speech recognition
  and understanding}}. IEEE Signal Processing Society.
\newblock


\bibitem[\protect\citeauthoryear{Qin, Carlini, Cottrell, Goodfellow, and
  Raffel}{Qin et~al\mbox{.}}{2019}]%
        {yaoqinicml}
\bibfield{author}{\bibinfo{person}{Yao Qin}, \bibinfo{person}{Nicholas
  Carlini}, \bibinfo{person}{Garrison Cottrell}, \bibinfo{person}{Ian
  Goodfellow}, {and} \bibinfo{person}{Colin Raffel}.}
  \bibinfo{year}{2019}\natexlab{}.
\newblock \showarticletitle{Imperceptible, Robust, and Targeted Adversarial
  Examples for Automatic Speech Recognition}. In
  \bibinfo{booktitle}{\emph{Proceedings of the 36th International Conference on
  Machine Learning}} \emph{(\bibinfo{series}{Proceedings of Machine Learning
  Research}, Vol.~\bibinfo{volume}{97})},
  \bibfield{editor}{\bibinfo{person}{Kamalika Chaudhuri} {and}
  \bibinfo{person}{Ruslan Salakhutdinov}} (Eds.). \bibinfo{publisher}{PMLR},
  \bibinfo{pages}{5231--5240}.
\newblock
\urldef\tempurl%
\url{http://proceedings.mlr.press/v97/qin19a.html}
\showURL{%
\tempurl}


\bibitem[\protect\citeauthoryear{Rabiner, Schafer, et~al\mbox{.}}{Rabiner
  et~al\mbox{.}}{1978}]%
        {FFT}
\bibfield{author}{\bibinfo{person}{Lawrence~R Rabiner},
  \bibinfo{person}{Ronald~W Schafer}, {et~al\mbox{.}}}
  \bibinfo{year}{1978}\natexlab{}.
\newblock \bibinfo{booktitle}{\emph{Digital processing of speech signals}}.
\newblock \bibinfo{publisher}{Prentice-hall}.
\newblock


\bibitem[\protect\citeauthoryear{Sch\"{o}nherr, Eisenhofer, Zeiler, Holz, and
  Kolossa}{Sch\"{o}nherr et~al\mbox{.}}{2020}]%
        {20imperio}
\bibfield{author}{\bibinfo{person}{Lea Sch\"{o}nherr},
  \bibinfo{person}{Thorsten Eisenhofer}, \bibinfo{person}{Steffen Zeiler},
  \bibinfo{person}{Thorsten Holz}, {and} \bibinfo{person}{Dorothea Kolossa}.}
  \bibinfo{year}{2020}\natexlab{}.
\newblock \showarticletitle{Imperio: Robust Over-the-Air Adversarial Examples
  for Automatic Speech Recognition Systems}. In
  \bibinfo{booktitle}{\emph{Annual Computer Security Applications Conference}}
  (Austin, USA) \emph{(\bibinfo{series}{ACSAC '20})}.
  \bibinfo{publisher}{Association for Computing Machinery},
  \bibinfo{address}{New York, NY, USA}, \bibinfo{pages}{843–855}.
\newblock
\showISBNx{9781450388580}
\urldef\tempurl%
\url{https://doi.org/10.1145/3427228.3427276}
\showDOI{\tempurl}


\bibitem[\protect\citeauthoryear{Sch{\"o}nherr, Kohls, Zeiler, Holz, and
  Kolossa}{Sch{\"o}nherr et~al\mbox{.}}{2018}]%
        {schonherr2018adversarial}
\bibfield{author}{\bibinfo{person}{Lea Sch{\"o}nherr},
  \bibinfo{person}{Katharina Kohls}, \bibinfo{person}{Steffen Zeiler},
  \bibinfo{person}{Thorsten Holz}, {and} \bibinfo{person}{Dorothea Kolossa}.}
  \bibinfo{year}{2018}\natexlab{}.
\newblock \showarticletitle{Adversarial attacks against automatic speech
  recognition systems via psychoacoustic hiding}.
\newblock \bibinfo{journal}{\emph{arXiv preprint arXiv:1808.05665}}
  (\bibinfo{year}{2018}).
\newblock


\bibitem[\protect\citeauthoryear{Shen and {et. al.}}{Shen and {et.
  al.}}{2019}]%
        {lingvo}
\bibfield{author}{\bibinfo{person}{Jonathan Shen} {and} \bibinfo{person}{{et.
  al.}}} \bibinfo{year}{2019}\natexlab{}.
\newblock \showarticletitle{Lingvo: a Modular and Scalable Framework for
  Sequence-to-Sequence Modeling}.
\newblock \bibinfo{journal}{\emph{CoRR}}  \bibinfo{volume}{abs/1902.08295}
  (\bibinfo{year}{2019}).
\newblock
\showeprint[arxiv]{1902.08295}
\urldef\tempurl%
\url{http://arxiv.org/abs/1902.08295}
\showURL{%
\tempurl}


\bibitem[\protect\citeauthoryear{Taori, Kamsetty, Chu, and Vemuri}{Taori
  et~al\mbox{.}}{2019}]%
        {blackboxSPW19}
\bibfield{author}{\bibinfo{person}{Rohan Taori}, \bibinfo{person}{Amog
  Kamsetty}, \bibinfo{person}{Brenton Chu}, {and} \bibinfo{person}{Nikita
  Vemuri}.} \bibinfo{year}{2019}\natexlab{}.
\newblock \showarticletitle{Targeted Adversarial Examples for Black Box Audio
  Systems}. In \bibinfo{booktitle}{\emph{2019 IEEE Security and Privacy
  Workshops (SPW)}}. \bibinfo{pages}{15--20}.
\newblock
\urldef\tempurl%
\url{https://doi.org/10.1109/SPW.2019.00016}
\showDOI{\tempurl}


\bibitem[\protect\citeauthoryear{Wang, Sun, Shan, Hou, Xie, Li, and Lei}{Wang
  et~al\mbox{.}}{2019}]%
        {adv_training2}
\bibfield{author}{\bibinfo{person}{Xiong Wang}, \bibinfo{person}{Sining Sun},
  \bibinfo{person}{Changhao Shan}, \bibinfo{person}{Jingyong Hou},
  \bibinfo{person}{Lei Xie}, \bibinfo{person}{Shen Li}, {and}
  \bibinfo{person}{Xin Lei}.} \bibinfo{year}{2019}\natexlab{}.
\newblock \showarticletitle{Adversarial Examples for Improving End-to-end
  Attention-based Small-footprint Keyword Spotting}. In
  \bibinfo{booktitle}{\emph{{IEEE} International Conference on Acoustics,
  Speech and Signal Processing, {ICASSP} 2019, Brighton, United Kingdom, May
  12-17, 2019}}. \bibinfo{publisher}{{IEEE}}, \bibinfo{pages}{6366--6370}.
\newblock
\urldef\tempurl%
\url{https://doi.org/10.1109/ICASSP.2019.8683479}
\showDOI{\tempurl}


\bibitem[\protect\citeauthoryear{Yakura and Sakuma}{Yakura and Sakuma}{2019}]%
        {hiromu}
\bibfield{author}{\bibinfo{person}{Hiromu Yakura} {and} \bibinfo{person}{Jun
  Sakuma}.} \bibinfo{year}{2019}\natexlab{}.
\newblock \showarticletitle{Robust Audio Adversarial Example for a Physical
  Attack}. In \bibinfo{booktitle}{\emph{Proceedings of the Twenty-Eighth
  International Joint Conference on Artificial Intelligence, {IJCAI-19}}}.
  \bibinfo{publisher}{International Joint Conferences on Artificial
  Intelligence Organization}, \bibinfo{pages}{5334--5341}.
\newblock
\urldef\tempurl%
\url{https://doi.org/10.24963/ijcai.2019/741}
\showDOI{\tempurl}


\bibitem[\protect\citeauthoryear{Yang, Li, Chen, and Song}{Yang
  et~al\mbox{.}}{2018}]%
        {yang2018characterizing}
\bibfield{author}{\bibinfo{person}{Zhuolin Yang}, \bibinfo{person}{Bo Li},
  \bibinfo{person}{Pin-Yu Chen}, {and} \bibinfo{person}{Dawn Song}.}
  \bibinfo{year}{2018}\natexlab{}.
\newblock \showarticletitle{Characterizing Audio Adversarial Examples Using
  Temporal Dependency}. In \bibinfo{booktitle}{\emph{International Conference
  on Learning Representations}}.
\newblock


\bibitem[\protect\citeauthoryear{Yuan, Chen, Zhao, Long, Liu, Chen, Zhang,
  Huang, Wang, and Gunter}{Yuan et~al\mbox{.}}{2018}]%
        {yuan2018commandersong}
\bibfield{author}{\bibinfo{person}{Xuejing Yuan}, \bibinfo{person}{Yuxuan
  Chen}, \bibinfo{person}{Yue Zhao}, \bibinfo{person}{Yunhui Long},
  \bibinfo{person}{Xiaokang Liu}, \bibinfo{person}{Kai Chen},
  \bibinfo{person}{Shengzhi Zhang}, \bibinfo{person}{Heqing Huang},
  \bibinfo{person}{Xiaofeng Wang}, {and} \bibinfo{person}{Carl~A Gunter}.}
  \bibinfo{year}{2018}\natexlab{}.
\newblock \showarticletitle{Commandersong: A systematic approach for practical
  adversarial voice recognition}. In \bibinfo{booktitle}{\emph{27th
  $\{$USENIX$\}$ Security Symposium ($\{$USENIX$\}$ Security 18)}}.
  \bibinfo{pages}{49--64}.
\newblock


\end{thebibliography}

\end{document}